\documentclass{article} % For LaTeX2e

\makeatletter
\def\input@path{{iclr_template/}}
\makeatother
\usepackage{preprint,times}

\setcitestyle{numbers,square} % cites becomes numbers
\usepackage{amsmath}
\usepackage{amssymb}
\usepackage{enumitem}

\usepackage{url}

\usepackage{booktabs}
\usepackage{wrapfig}
\usepackage{multirow}
\usepackage{array}
\usepackage{tabularx}

\usepackage{graphicx}
\usepackage{adjustbox}
\usepackage{xcolor}
\usepackage{soul}

\usepackage[
    caption=false,
    font=footnotesize,
    labelfont=sf,
    textfont=sf
]{subfig}

\usepackage{float}
\usepackage{placeins}
\usepackage{needspace}

\usepackage{tikz}
\usepackage{pgfplots}
\usepgfplotslibrary{groupplots}
\pgfplotsset{compat=1.18}

\usetikzlibrary{
    positioning,
    calc,
    arrows.meta,
    shapes.symbols,
    fit,
    backgrounds,
    shapes.multipart
}

\tikzset{
    clapcloud/.style={
        draw=black!75,
        fill=black!6,
        line width=0.85pt,
        line join=round
    },
    task/.style={
        rounded corners=3pt,
        draw=black!70,
        fill=white,
        line width=0.65pt,
        text width=3.72cm,
        minimum height=3.25cm,
        align=center,
        inner sep=5pt
    },
    evaluation/.style={
        rounded corners=3pt,
        draw=black!60,
        fill=black!4,
        dashed,
        line width=0.75pt,
        text width=16.55cm,
        minimum height=1.65cm,
        align=center,
        inner sep=6pt
    },
    flow/.style={
        -{Latex[length=3.0mm,width=2.0mm]},
        draw=black!85,
        line width=1.35pt
    },
    bus/.style={
        draw=black!60,
        line width=0.75pt
    },
    auxiliary/.style={
        -{Latex[length=2.6mm,width=1.7mm]},
        draw=black!65,
        dashed,
        line width=1.0pt
    }
}

\newcommand{\DownstreamModelCell}[2]{%
  \begin{tabular}[c]{@{}l@{}}
    {\footnotesize\bfseries #1}\\[-1pt]
    {\footnotesize\bfseries (#2)}
  \end{tabular}%
}

\definecolor{ConceptBenign}{HTML}{1B5E20}
\colorlet{ConceptHarm}{red}

\usepackage{hyperref}
\hypersetup{colorlinks=true,linkcolor=blue,citecolor={red!70!black},urlcolor=blue}

\soulregister{\cite}{7}
\soulregister{\citep}{7}
\soulregister{\citet}{7}
\soulregister{\ref}{7}
\soulregister{\eqref}{7}

\renewenvironment{abstract}
    {\vskip.075in
     \centerline{\large\bfseries Abstract}
     \vspace{0.5ex}
     \begin{quote}}
    {\par\end{quote}\vskip 1ex}

\title{Concept-Based Adversarial Audio: \\ Query-Free Attacks via CLAP Dependencies}
\author{
    \textbf{Rasmus Torp}\thanks{Equal contribution.}
    \quad
    \textbf{Manzi Fabrice Niyigaba}\footnotemark[1]
    \quad
    \textbf{Adam Breuer}
    \\[5pt]
    Dartmouth College
}

\begin{document}

\maketitle

\begin{abstract}
In this paper, we introduce query-free attacks that manipulate models across the diverse universe of audio AI by exploiting their common CLAP dependency, without accessing or querying them. Public Contrastive Language–Audio Pretraining (CLAP) models are widely reused as input audio encoders across otherwise independent systems for audio classification, generation, retrieval, captioning, and question answering. We study a query-free threat model where an adversary has white-box access only to the public CLAP checkpoints, but no access or queries to the target model and no knowledge of its task, architecture, parameters, or training data.

We study two classes of attacks, both optimized entirely offline against the public checkpoints. First, we study conventional attacks that apply a small norm-bounded perturbation to the waveform to change CLAP's output using ensemble PGD, then test transfer to the unseen downstream system. Second, we introduce the \emph{first concept-based adversarial audio attack}. These attacks generate replacement audio that stays within the same semantic concept as the original (e.g., sound of a slightly different gunshot; meow of a slightly different kitten) while being incorrectly encoded by CLAP.

We show that both attacks transfer downstream, and concept attacks in particular transfer even to peripheral CLAP-family-dependent systems that share no public CLAP checkpoint parameters. These attacks can continue to survive under downsampling, low-pass filtering, MP3 compression, \emph{and over-the-air playback and re-recording} on cheap consumer speakers/microphones outdoors amid heavy ambient noise in Hollywood’s recording district. 
%
% \textbf{\textcolor{red}{CONCEPT TRANSFERS BETTER}}
%
%Our results identify shared public encoders as a common-mode vulnerability through which attacks can propagate across downstream systems without target knowledge, access, or queries.
\end{abstract}

\section{Introduction}
\label{sec:intro}
Suppose an adversary wants to manipulate the decisions of an audio AI system that it cannot inspect or query and knows almost nothing about. For example, the attacker may want benign audio to be interpreted as gunfire or screaming, with the goal of triggering a false-positive moderation response; alternatively, the attacker may want harmful audio, such as audio containing violence or mature themes, to appear benign and evade moderation or prevent a response from being dispatched. % Required once in the document preamble:
% \usetikzlibrary{fit,backgrounds,shapes.multipart,shapes.symbols,calc,arrows.meta}
%
% Kept here too so this file can be dropped into a project whose preamble
% only loaded the older TikZ libraries.
\usetikzlibrary{shapes.symbols,calc,arrows.meta}

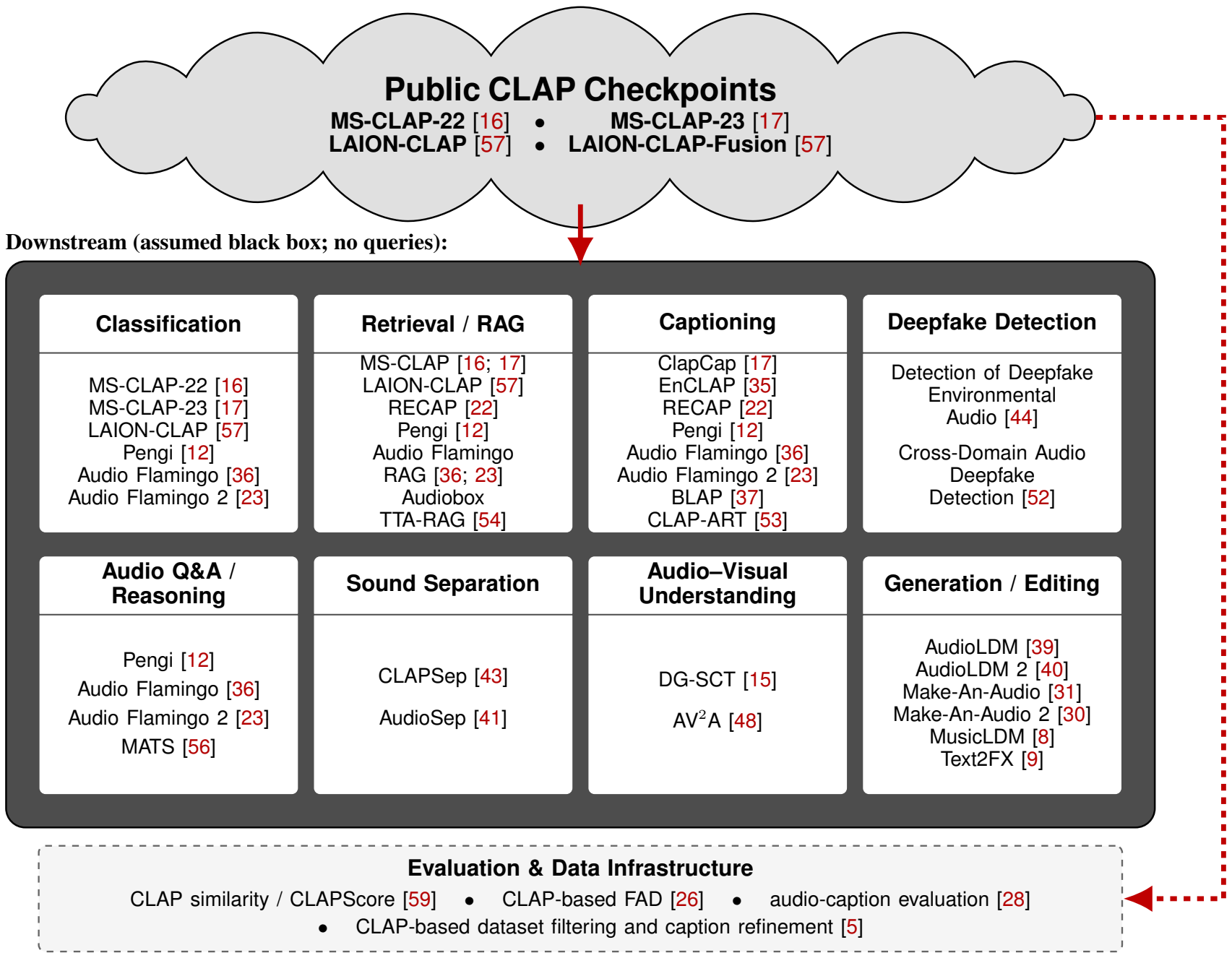
\begin{figure*}[!t]
\centering

\begin{adjustbox}{max width=0.8\textwidth}
\begin{tikzpicture}[font=\sffamily]

\tikzset{
    clapcloud/.style={
        cloud,
        cloud puffs=18,
        cloud puff arc=114,
        aspect=8.2,
        draw=black,
        line width=0.8pt,
        fill=black!12,
        inner xsep=4mm,
        inner ysep=0pt,
        minimum width=12.6cm,
        minimum height=7mm,
        text width=10.6cm,
        align=center
    },
    task/.append style={
        rectangle split,
        rectangle split parts=2,
        rectangle split draw splits=true,
        rectangle split part fill={black!25,white},
        rectangle split part align={center,center},
        inner ysep=0pt
    },
    flow/.style={
        -{Latex[length=4.6mm,width=4.2mm]},
        draw=red!75!black,
        line width=2.2pt
    },
    auxiliary/.style={
        -{Latex[length=4.6mm,width=4.2mm]},
        draw=red!75!black,
        dashed,
        line width=2.2pt
    }
}

% A fixed-height title area keeps every task heading aligned.
\newcommand{\tasktitle}[1]{%
    \parbox[c][9mm][c]{3.6cm}{%
        \centering\bfseries\normalsize #1%
    }%
}

% Together with the 8 mm title, this gives each task box a fixed
% internal height of 50 mm.
\newcommand{\taskbody}[1]{%
    \parbox[c][28mm][c]{3.2cm}{%
        \centering\footnotesize #1%
    }%
}

% ------------------------------------------------------------------
% First row of downstream domains
% ------------------------------------------------------------------

\node[task] (classification) at (-6.45,1.60) {
    \tasktitle{Classification}
    \nodepart{second}
    \taskbody{
        MS-CLAP-22~\cite{elizalde_clap_2023}\\
        MS-CLAP-23~\cite{elizalde_natural_2024}\\
        LAION-CLAP~\cite{wu_large-scale_2024}\\
        Pengi~\cite{deshmukh_pengi_2024}\\
        Audio Flamingo~\cite{kong_audio_2024}\\
        Audio Flamingo 2~\cite{ghosh_audio_2025}
    }
};

\node[task] (retrieval) at (-2.15,1.60) {
    \tasktitle{Retrieval / RAG}
    \nodepart{second}
    \taskbody{
        MS-CLAP%
            ~\cite{elizalde_clap_2023,elizalde_natural_2024}\\
        LAION-CLAP~\cite{wu_large-scale_2024}\\
        RECAP~\cite{ghosh_recap_2024}\\
        Pengi~\cite{deshmukh_pengi_2024}\\
        Audio Flamingo RAG~\cite{kong_audio_2024,ghosh_audio_2025}\\
        Audiobox TTA-RAG~\cite{vyas_audiobox_2023}
    }
};

\node[task] (captioning) at (2.15,1.60) {
    \tasktitle{Captioning}
    \nodepart{second}
    \taskbody{
        ClapCap~\cite{elizalde_natural_2024}\\
        EnCLAP~\cite{kim_enclap_2024}\\
        RECAP~\cite{ghosh_recap_2024}\\
        Pengi~\cite{deshmukh_pengi_2024}\\
        Audio Flamingo~\cite{kong_audio_2024}\\
        Audio Flamingo 2~\cite{ghosh_audio_2025}\\
        BLAP~\cite{lanzendorfer_blap_nodate}\\
        CLAP-ART~\cite{takeuchi_clap-art_2025}
    }
};

\node[task] (deepfake) at (6.45,1.60) {
    \tasktitle{Deepfake Detection}
    \nodepart{second}
    \taskbody{
        Detection of Deepfake\\
        Environmental Audio~\cite{ouajdi_detection_2024}\\[2mm]
        Cross-Domain Audio Deepfake Detection~\cite{sriram_cross-domain_2027}\\[2mm]
    }
};

% ------------------------------------------------------------------
% Second row of downstream domains
% ------------------------------------------------------------------

\node[task] (aqa) at (-6.45,-2.50) {
    \tasktitle{Audio Q\&A / Reasoning}
    \nodepart{second}
    \taskbody{
        Pengi~\cite{deshmukh_pengi_2024}\\[1.0mm]
        Audio Flamingo~\cite{kong_audio_2024}\\[1.0mm]
        Audio Flamingo 2~\cite{ghosh_audio_2025}\\[1.0mm]
        MATS~\cite{wang_mats_2025}
    }
};

\node[task] (separation) at (-2.15,-2.50) {
    \tasktitle{Sound Separation}
    \nodepart{second}
    \taskbody{
        CLAPSep~\cite{ma_clapsep_2024}\\[3mm]
        AudioSep~\cite{liu_separate_2025}\\[2mm]
    }
};

\node[task] (audiovisual) at (2.15,-2.50) {
    \tasktitle{Audio--Visual\\Understanding}
    \nodepart{second}
    \taskbody{
        DG-SCT~\cite{duan_cross-modal_2023}\\[3mm]
        AV$^2$A~\cite{shaar_adapting_nodate}\\[1.0mm]
    }
};

\node[task] (generation) at (6.45,-2.50) {
    \tasktitle{Generation / Editing}
    \nodepart{second}
    \taskbody{
        AudioLDM~\cite{liu_audioldm_2023}\\
        AudioLDM 2~\cite{liu_audioldm_2024}\\
        Make-An-Audio~\cite{huang_make--audio_2023-1}\\
        Make-An-Audio 2~\cite{huang_make--audio_2023}\\
        MusicLDM~\cite{chen_musicldm_2024}\\
        Text2FX~\cite{chu_text2fx_2025}
    }
};

% ------------------------------------------------------------------
% Black-box region fitted evenly around the downstream task nodes.
% ------------------------------------------------------------------

\begin{scope}[on background layer]
\node[
    fit=(classification)
        (retrieval)
        (captioning)
        (deepfake)
        (aqa)
        (separation)
        (audiovisual)
        (generation),
    inner xsep=5mm,
    inner ysep=5mm,
    draw=black!,
    % dashed,
    line width=0.8pt,
    rounded corners=3mm,
    fill=black!70
] (blackbox) {};
\end{scope}

\node[
    anchor=south west,
    font=\bfseries\small,
    inner sep=0pt,
    yshift=1mm
] at (blackbox.north west) {\normalsize Downstream (assumed black box; no queries):};

% ------------------------------------------------------------------
% Attack optimization over the public checkpoint ensemble
% ------------------------------------------------------------------

\node[clapcloud, anchor=south] (attack) at ([yshift=5mm]blackbox.north) {
    \textbf{\Large Public CLAP Checkpoints}\\[0.4mm]
    \renewcommand{\arraystretch}{0.9}%
    \begin{tabular}{@{}c@{\quad$\bullet$\quad}c@{}}
        \textbf{MS-CLAP-22}~\cite{elizalde_clap_2023}
            & \textbf{MS-CLAP-23}~\cite{elizalde_natural_2024} \\
        \textbf{LAION-CLAP}~\cite{wu_large-scale_2024}
            & \textbf{LAION-CLAP-Fusion}~\cite{wu_large-scale_2024}
    \end{tabular}
};

\draw[flow]
    ([yshift=3.8mm]attack.south) -- ([yshift=-1mm]blackbox.north);

% ------------------------------------------------------------------
% Evaluation and research infrastructure (separate access path)
% ------------------------------------------------------------------

\node[evaluation] (evaluation) at (0,-6) {
    \textbf{\normalsize Evaluation \& Data Infrastructure}\\[1.2mm]
    \footnotesize
    CLAP similarity / CLAPScore%
        ~\cite{xiao_reference-free_2025}
    \quad $\bullet$ \quad
    CLAP-based FAD%
        ~\cite{gui_adapting_2024}
    \quad $\bullet$ \quad
    audio-caption evaluation%
        ~\cite{guo_brace_2026}\\[0.5mm]
    \quad $\bullet$ \quad
    CLAP-based dataset filtering and caption refinement%
        ~\cite{bai_audiosetcaps_2025}
};

\coordinate (evalroute) at ([xshift=20mm]attack.east);

\draw[auxiliary]
    (attack.east) -- (evalroute) |- (evaluation.east);

\end{tikzpicture}
\end{adjustbox}

\caption{
Diverse downstream audio systems contain a CLAP-family dependency.
}
\label{fig:clap-ecosystem}

\end{figure*}

In this paper, we show that it is enough that the attacker suspects that the unknown system depends somewhere on public Contrastive Language--Audio Pretraining (CLAP). This is plausible because CLAP has achieved widespread popularity across diverse audio AI domains, including in security-critical domains such as audio deepfake detection (e.g.  \cite{ouajdi_detection_2024,sriram_cross-domain_2027}). Public CLAP model checkpoints are ubiquitously reused as audio encoders and initialization points across otherwise independent systems for audio \cite{elizalde_clap_2023,elizalde_natural_2024,wu_large-scale_2024}. Specifically, CLAP representations are used for audio classification and general-purpose audio-language modeling (e.g.  \cite{elizalde_clap_2023,elizalde_natural_2024,wu_large-scale_2024,deshmukh_pengi_2024,kong_audio_2024,ghosh_audio_2025}); retrieval and retrieval-augmented generation (e.g.  \cite{ghosh_recap_2024,deshmukh_pengi_2024,kong_audio_2024,ghosh_audio_2025,vyas_audiobox_2023}; captioning (e.g.  \cite{elizalde_natural_2024,kim_enclap_2024,ghosh_recap_2024,deshmukh_pengi_2024,kong_audio_2024,ghosh_audio_2025,lanzendorfer_blap_nodate,takeuchi_clap-art_2025}); audio question answering and reasoning (e.g. \cite{deshmukh_pengi_2024,kong_audio_2024,ghosh_audio_2025,wang_mats_2025}); sound separation (e.g.  \cite{ma_clapsep_2024,liu_separate_2025}); audio--visual understanding (e.g. \cite{duan_cross-modal_2023,shaar_adapting_nodate}); and audio generation and editing (e.g.  \cite{liu_audioldm_2023,liu_audioldm_2024,huang_make--audio_2023,huang_make--audio_2023-1,chen_musicldm_2024,chu_text2fx_2025}). CLAP representations are also used to evaluate other audio systems, compute CLAP-based quality metrics, filter datasets, and refine training captions (e.g. \cite{xiao_reference-free_2025,gui_adapting_2024,guo_brace_2026,bai_audiosetcaps_2025}).
Fig.~\ref{fig:clap-ecosystem} maps a small subset of these dependencies; audio systems developed by different groups, for different tasks, with different architectures and training procedures repeatedly depend on the same small family of public CLAP representations.

Armed with nothing but this reasonable dependency suspicion, the attacker can exploit a public attack surface in the target: its CLAP dependency. While the unseen downstream system remains inaccessible, the attacker can download the public CLAP dependency checkpoints and construct white-box attacks against them entirely offline. We consider whether that is enough to corrupt the decisions of an unseen, unqueried downstream system.

To study this problem, we consider a deliberately restrictive, query-free threat model in which the attacker constructs attacks using only public CLAP checkpoints and receives no feedback from the downstream target. We consider two attack-construction mechanisms, \emph{Perturbation} \& \emph{Synthesis}, and two downstream adversarial goals, \emph{bypass} \& \emph{trigger}. \S~\ref{sec:threat-model} formalizes this threat model.

Under this threat model, we make three core contributions:

\noindent\textbf{I: Public CLAP encoders constitute a shared attack surface across downstream audio AI.}

We identify and empirically study a shared attack surface arising from the widespread reuse of public CLAP encoders across otherwise independently developed audio systems. Under a query-free threat model, we demonstrate that conventional waveform perturbation attacks optimized exclusively against public CLAP checkpoints transfer to previously unseen downstream systems without knowledge of their architecture, parameters, or task. Our evaluation spans five downstream systems, four datasets, and four task directions, including systems that incorporate CLAP as a frozen component, fine-tune CLAP-pretrained encoders, or employ independently trained CLAP-family representations.

% \noindent\textbf{Contribution I: Public CLAP encoders constitute a shared attack surface across downstream audio AI.}
% For \emph{Perturbation}, we begin with a standard tool: PGD on the waveform. Specifically, we optimize \textsc{clapPGD} jointly over four public CLAP checkpoints, following prior evidence that ensembles can improve audio transfer \cite{guo_towards_2023}. However, conventional PGD attacks optimize against the target model or against a target-specific surrogate to improve transfer (e.g \cite{gubri_lgv_2022}). Here, the attacker has neither; it sees only public CLAP checkpoints and has no target system knowledge with which to make a full surrogate. Nor is the downstream target simply CLAP plus a head: for example, an EnCLAP target combines frozen CLAP embeddings with a separate neural audio-codec representation, AF1 places a frozen CLAP-derived encoder inside a much larger audio-language model, and Pengi substantially fine-tunes its CLAP-pretrained encoder. Thus, there is good reason to expect transfer to fail. Indeed, representation-space attacks can transfer poorly when source and target geometries are misaligned \cite{gupta_understanding_2025}, and audio transfer has historically been especially weak across heterogeneous models and for targeted attacks (e.g \cite{abdullah_demystifying_nodate,hussain_soundbreak_2026}). 

% Surprisingly, we show that our PGD attacks on public CLAP checkpoints can transfer to unseen downstream systems that peripherally depend on them.

\noindent\textbf{II: Concept-based adversarial audio constitutes a distinct attack vector.}

We introduce \textsc{ConceptAttack}, to our knowledge the first concept-based adversarial attack for audio.
Rather than restricting the adversary to a norm-bounded neighborhood of a particular waveform, \textsc{ConceptAttack} exploits the semantic freedom available when the adversary can author the submitted audio.
Building on recent concept-based attacks in vision \cite{zhang_concept-based_2026}, we develop a synthesis-based attack that combines adversarially guided audio generation with automated source-concept preservation criteria based on independent audio representation models.

While conventional perturbation attacks frequently match or outperform \textsc{ConceptAttack} on the attacked public checkpoints, \textsc{ConceptAttack} exhibits substantially stronger transfer to more distant downstream systems, including independently trained CLAP-family encoders. For example, on AF-CLAP, which shares no attacked checkpoint parameters, \textsc{ConceptAttack} achieves $52.0\%$ targeted ASR@10 for AudioCaps text-to-audio retrieval, compared with $6.1\%$ for \textsc{clapPGD} and $3.8\%$ for Adversarial Illusions.

% \noindent\textbf{Contribution II: The first concept-based attacks for audio.}
% For \emph{Synthesis}, the threat model permits the attacker to author the submitted audio. We use this additional freedom to construct novel attacks constituting a distinct threat-vector, and enabling potentially stronger robustness and transfer to distant CLAP-derived systems. Concept-based attacks have just emerged in vision: Zhang et al. introduced the formulation at ICLR 2026 \cite{zhang_concept-based_2026}. In this paper, we introduce the first \emph{concept-based adversarial audio}. We guide audio generation to produce replacement recordings that remain valid instances of the source concept while CLAP encodes them toward an attacker-chosen target concept. 

\noindent\textbf{III: Concept-based adversarial audio survives:}

\begin{enumerate}[label=(\alph*)]
    \item \textbf{Model adaptation.}
    \textsc{ConceptAttack} remains effective against downstream
    systems that fine-tune and adapt CLAP-pretrained encoders.
    We further demonstrate through controlled fine-tuning
    experiments that standard fine-tuning of CLAP does not
    eliminate the attacks.

    \item \textbf{Aggressive signal processing.}
    \textsc{ConceptAttack} exhibits substantially greater robustness
    than conventional waveform perturbation attacks to MP3
    compression, resampling, low-pass filtering, and additive noise.
    Even after round-trip resampling to $8$ kHz,
    \textsc{ConceptAttack} achieves approximately $74\%$ targeted
    ASR on LAION-CLAP across both classification datasets.

    \item \textbf{Physical playback and re-recording.}
    \textsc{ConceptAttack} remains effective after over-the-air
    playback and re-recording using consumer-grade equipment
    in noisy outdoor environments.
    Even when re-recorded using an iPhone's built-in microphone,
    \textsc{ConceptAttack} achieves $18$--$19\%$ targeted ASR
    against Audio Flamingo 2 (AF2) \cite{ghosh_audio_2025},
    whose audio encoder shares no parameters with the attacked
    public CLAP checkpoints.
\end{enumerate}

\noindent\textbf{Security implications.}

Together, these findings demonstrate that the \emph{widespread reuse of foundation audio encoders creates a shared security vulnerability across otherwise independently developed audio systems.} Concept-based adversarial audio exposes the extent of this vulnerability by transferring beyond shared model parameters and surviving digital and physical transmission.

\section{Threat Model}
\label{sec:threat-model}
The adversary assumes only that the (potentially sophisticated) target model depends on CLAP. The downstream target model is itself inaccessible during attack construction. The adversary instead constructs the submitted audio entirely offline against public CLAP checkpoints, then supplies it via the target's ordinary input channel. Table~\ref{tab:attack-framework} summarizes the two construction mechanisms permitted by the threat model and the two downstream goals.

\begin{table*}[t]
\centering
\caption{Threat-model mechanisms, their algorithmic realizations, and the two downstream goals we consider. The mechanism specifies the freedom available to the adversary; the algorithm is the corresponding attack we develop.}
\label{tab:attack-framework}
\small
\setlength{\tabcolsep}{5pt}
\setlength{\extrarowheight}{1pt}
\renewcommand{\arraystretch}{1.08}
\begin{tabularx}{\textwidth}{
    @{}
    l
    l
    >{\raggedright\arraybackslash}X
    >{\raggedright\arraybackslash}X
    @{} }
\toprule
& & \multicolumn{2}{c}{\textbf{Adversarial goal}} \\
\cmidrule(lr){3-4}
\textbf{Mechanism}
& \textbf{Algorithm}
& \multicolumn{1}{c}{\textbf{Bypass}}
& \multicolumn{1}{c}{\textbf{Trigger}} \\
\midrule

\textbf{Perturbation}
&
Ensemble PGD
&
Modify source audio (e.g., harmful) so the downstream system redirects it away from the source concept, with the goal of bypassing a downstream response. % AVOID USING THE WORD LISTENER
&
Modify source audio (e.g., benign) so the downstream system redirects it toward an attacker-chosen target concept such as gunfire or screaming, with the goal of triggering a downstream response. % AVOID USING THE WORD LISTENER
\\

\addlinespace[0.22em]
\cmidrule(lr){1-4}
\addlinespace[0.22em]

\textbf{Synthesis}
&
Concept-based
&
Author semantically valid harmful audio that the system instead interprets as some benign class.
&
Author semantically benign-class audio that the system interprets as an attacker-chosen (e.g., harmful) class.
\\

\bottomrule
\end{tabularx}
\end{table*}

\medskip\noindent\textbf{Deployment setting and adversarial goals.}
The consequence of a corrupted audio representation depends on what the downstream system does with it. Our motivating cases are systems in which an audio interpretation informs or triggers a response, including content moderation, acoustic event detection, and security-critical applications such as audio deepfake detection.

We consider two deployment-level goals. In \emph{bypass}, the adversary seeks to change the system's interpretation to preclude a downstream response it wants to avoid. In \emph{trigger}, the adversary seeks to change the system's interpretation to cause a downstream response of its choosing. Whether a particular semantic redirection realizes either goal depends on the downstream system and policy, which are unknown to the adversary.

\emph{Bypass} and \emph{trigger} describe the adversary's security goals, not the optimization losses used to construct an attack. Our experiments do not assume a particular downstream policy. Instead, they measure whether attacks constructed only against public CLAP checkpoints change the outputs of previously unseen downstream systems under the targeted and untargeted criteria defined in \S \ref{sec:methods}.

\medskip\noindent\textbf{Formal setting.}
Let $f_1,\ldots,f_m$ denote the public CLAP audio encoders available to the adversary, and let $g$ denote the unknown audio representation used by the downstream target. For targets in scope, $g$ inherits from CLAP, but the relationship between $g$ and the public encoders $\{f_j\}_{j=1}^m$ is unknown to the adversary. The adversary constructs the attack using only the public $f_j$; $g$ is never evaluated during attack construction.

Attack validity also requires semantic preservation. The submitted audio must remain a valid instance of its source concept; simply corrupting the audio until the source concept is lost does not count. The Methods section specifies how we evaluate this condition for Perturbation and Synthesis, respectively.

\medskip\noindent\textbf{Perturbation and synthesis.}
In a \emph{perturbation} attack, the adversary begins with a particular existing recording $x_s$ and must preserve that recording while modifying it. The submitted audio takes the form $x_{\mathrm{adv}}=x_s+\delta$, subject to a small perturbation constraint. This mechanism applies when the particular recording matters to the adversary---for example, when a specific clip must still be delivered. Proximity to $x_s$ is therefore an explicitly enforced constraint during attack construction. We instantiate Perturbation with ensemble PGD.

In a \emph{synthesis} attack, the adversary may author the submitted recording outright. No particular source waveform is privileged by the threat model. A different gunshot can be a perfectly ordinary gunshot; a different meow can be a perfectly ordinary meow. Instead, the constraint is semantic: the submitted audio must remain a valid realization of the source concept.

Under \emph{Synthesis}, there is no privileged reference waveform. A conventional constraint such as
\[
    \lVert x_{\mathrm{adv}}-x_s \rVert \leq \varepsilon
\]
defines similarity relative to one particular clean waveform $x_s$. That constraint is appropriate when the adversary must preserve $x_s$. Under Synthesis, the adversary authors the submitted audio file and the target receives only that file. Restricting the attack to an arbitrary waveform's $\varepsilon$-ball therefore discards degrees of freedom the threat model permits. We instantiate Synthesis with concept-based adversarial audio, which preserves the source concept rather than a particular waveform.  Section~\ref{sec:judges} describes how we enforce this semantic constraint during attack construction using independent audio models.

\medskip\noindent\textbf{Adversary's capabilities.}
The adversary has white-box access to publicly available CLAP checkpoints, including their parameters and gradients, and may use several public checkpoints jointly during offline attack construction. These checkpoints constitute the adversary's only target-specific model knowledge. The adversary may otherwise use non-target-specific public resources such as source audio, text descriptions, and public audio generators.

The downstream target is never queried during attack construction. The adversary does not know its task, architecture, parameters, training data, fine-tuned parameters, prediction head, prompts, preprocessing, inference procedure, particular CLAP ancestor, extent of downstream adaptation, or deployment scenario. The adversary receives no target output or other feedback. We use \emph{query-free} to mean that no downstream output or feedback is used to construct, select, or refine the attack.

% Keep this wrapped table close to its surrounding paragraphs.
\begingroup
\setlength{\intextsep}{3pt}
\begin{wraptable}{r}{0.58\textwidth}
\centering
\setlength{\abovecaptionskip}{0pt}
\caption{\raggedright Examples of attacks on unseen downstream systems.}
\label{tab:attack-audio-examples}
\footnotesize
\setlength{\tabcolsep}{2.5pt}
\renewcommand{\arraystretch}{1.05}
\begin{tabularx}{\linewidth}{@{}c>{\raggedright\arraybackslash}Xll@{}}
\toprule
\textbf{Attack}
& \textbf{Effect}
& \textbf{System}
& \textbf{Audio} \\
\midrule

\multirow[c]{6}{*}[-0.791pt]{\textsc{clapPGD}}
& \textcolor{ConceptBenign}{Snoring Man} $\rightarrow$\newline \textcolor{ConceptHarm}{Several Gunshots}
& EnCLAP, AF1
& \href{https://anonymous.4open.science/r/AdversarialAudio-1A78/assets/example_audios/clapPGD/SnoringMan_to_SeveralGunShots_AF1_and_Enclap.wav}{[\textsc{listen}]} \\
\addlinespace[5pt]

& \textcolor{ConceptHarm}{Police Sirens} $\rightarrow$\newline \textcolor{ConceptBenign}{Woman Talking}
& EnCLAP
& \href{https://anonymous.4open.science/r/AdversarialAudio-1A78/assets/example_audios/clapPGD/SirenAndMotorcycle_to_WomanTalkingSowingMachine_EnClap.wav}{[\textsc{listen}]} \\
\addlinespace[5pt]

& \textcolor{ConceptHarm}{Baby Crying} $\rightarrow$\newline \textcolor{ConceptBenign}{Birds and Frogs}
& EnCLAP
& \href{https://anonymous.4open.science/r/AdversarialAudio-1A78/assets/example_audios/clapPGD/CryingBaby_to_BirdsAndFrogs_EnClap.wav}{[\textsc{listen}]} \\

\midrule

\multirow[c]{3}{*}[-1.541pt]{ConceptAttack}
& \mbox{\textcolor{ConceptBenign}{Birds} $\rightarrow$ \textcolor{ConceptHarm}{Siren}}
& EnCLAP
& \href{https://anonymous.4open.science/r/AdversarialAudio-1A78/assets/example_audios/concepts/birds_to_siren_enclap/adversarial.wav}{[\textsc{listen}]} \\
\addlinespace[2pt]

& \mbox{\textcolor{ConceptHarm}{Gunshot} $\rightarrow$ \textcolor{ConceptBenign}{Hammer}}
& AF1
& \href{https://anonymous.4open.science/r/AdversarialAudio-1A78/assets/example_audios/concepts/gunshot_to_hammer_af1_pengi/adversarial.wav}{[\textsc{listen}]} \\
\addlinespace[2pt]

& \mbox{\textcolor{ConceptHarm}{Scream} $\rightarrow$ \textcolor{ConceptBenign}{Whistle}}
& AF2
& \href{https://anonymous.4open.science/r/AdversarialAudio-1A78/assets/example_audios/concepts/scream_to_whistle_af2/adversarial.wav}{[\textsc{listen}]} \\

\bottomrule
\end{tabularx}
\end{wraptable}

% \begin{table}[h]
% \centering
% \caption{Examples of attacks on unseen downstream systems.}
% \label{tab:attack-audio-examples}
% \footnotesize
% \begin{tabular}{llll}
% \toprule
% \textbf{Attack}
% & \textbf{Effect}
% & \textbf{Downstream system}
% & \textbf{Audio} \\
% \midrule
% \textsc{clapPGD}
% & \textcolor{ConceptBenign}{Snoring Man} $\rightarrow$ \textcolor{ConceptHarm}{Several Gunshots}
% & EnCLAP, AF1
% & \href{https://anonymous.4open.science/r/AdversarialAudio-1A78/assets/example_audios/clapPGD/SnoringMan_to_SeveralGunShots_AF1_and_Enclap.wav}{[\textsc{listen}]} \\
% \textsc{clapPGD}
% & \textcolor{ConceptHarm}{Police Sirens} $\rightarrow$ \textcolor{ConceptBenign}{Woman Talking}
% & EnCLAP
% & \href{https://anonymous.4open.science/r/AdversarialAudio-1A78/assets/example_audios/clapPGD/SirenAndMotorcycle_to_WomanTalkingSowingMachine_EnClap.wav}{[\textsc{listen}]} \\
% \textsc{clapPGD}
% & \textcolor{ConceptHarm}{Baby Crying} $\rightarrow$ \textcolor{ConceptBenign}{Birds and Frogs}
% & EnCLAP
% & \href{https://anonymous.4open.science/r/AdversarialAudio-1A78/assets/example_audios/clapPGD/CryingBaby_to_BirdsAndFrogs_EnClap.wav}{[\textsc{listen}]} \\

% AudioConcept
% & \textcolor{ConceptBenign}{Birds} $\rightarrow$ \textcolor{ConceptHarm}{Siren}
% & EnCLAP
% & \href{https://anonymous.4open.science/r/AdversarialAudio-1A78/assets/example_audios/concepts/birds_to_siren_enclap/adversarial.wav}{[\textsc{listen}]} \\

% AudioConcept
% & \textcolor{ConceptHarm}{Gunshot} $\rightarrow$ \textcolor{ConceptBenign}{Hammer}
% & AF1
% & \href{https://anonymous.4open.science/r/AdversarialAudio-1A78/assets/example_audios/concepts/gunshot_to_hammer_af1_pengi/adversarial.wav}{[\textsc{listen}]} \\

% AudioConcept
% & \textcolor{ConceptHarm}{Scream} $\rightarrow$ \textcolor{ConceptBenign}{Whistle}
% & AF2
% & \href{https://anonymous.4open.science/r/AdversarialAudio-1A78/assets/example_audios/concepts/scream_to_whistle_af2/adversarial.wav}{[\textsc{listen}]} \\

% \bottomrule
% \end{tabular}
% \end{table}

\noindent\textbf{Target systems and CLAP ancestry.}
We consider systems whose audio representations inherit from public CLAP models. A target may use a released CLAP checkpoint directly, consume frozen CLAP embeddings, fine-tune a CLAP-initialized encoder, or incorporate a CLAP-derived encoder into a larger audio or multimodal system. The adversary does not know which relationship holds for a given target. Table~\ref{tab:downstream-clap-lineage} records the relationships for the systems we evaluate.

The evaluated targets span different degrees of CLAP ancestry. At the more distant end, Audio Flamingo 2 uses a custom AF-CLAP checkpoint and staged adaptation rather than directly reusing one of our public surrogate checkpoints. We include it as the most distant CLAP-derived case in our evaluation.
\par\endgroup

\medskip\noindent\textbf{Attack delivery.}
Our primary setting is digital audio supplied through the target's ordinary input channel. This may be an uploaded audio file, an item inserted into a retrieval corpus, or audio supplied to a classification or captioning system. The target may apply compression, resampling, normalization, frequency filtering, or other transformations unknown to the adversary.

We additionally evaluate a physical-delivery extension in which the same adversarial audio is played through a loudspeaker in a noisy environment and re-recorded through a microphone. The attack is constructed before playback and receives no information about the environment or playback-and-recording chain.

\medskip\noindent\textbf{Scope.}
The adversary submits audio through an ordinary input channel. The adversary does not modify or retrain the target, poison its training data, compromise its serving infrastructure or delivery pipeline, modify its prompts, or attack a non-audio modality. Downstream developers may freeze, fine-tune, or otherwise adapt CLAP normally; the adversary does not tamper with that process. Systems unrelated to CLAP fall outside our scope. Our empirical evaluation focuses on general-purpose CLAP-family systems for environmental audio and audio--language tasks. We do not evaluate speech-specific systems such as automatic speech recognition or music-specific foundation models; whether the same dependency attack transfers to those domains remains open.

Our moderation, detection, and automated-response examples motivate the security consequence of the threat model. We do not attack a deployed commercial moderation or safety system and make no claim about any particular production system. Our empirical targets are the research systems summarized in Table~\ref{tab:downstream-clap-lineage}; they test whether adversarial manipulation of public CLAP representations survives downstream reuse and adaptation.

\section{Methods}
\label{sec:methods}
We instantiate the two attack-construction mechanisms defined in Section~\ref{sec:threat-model}. For \emph{Perturbation}, \textsc{clapPGD} optimizes a bounded waveform perturbation against only an ensemble of public CLAP checkpoints, then attempts to transfer to the unseen downstream system. For \emph{Synthesis}, we introduce \textsc{ConceptAttack}, the first concept-based adversarial audio attack, which searches over newly generated realizations of the source concept for examples that CLAP encodes as an attacker-chosen target concept.

\begin{figure*}[t]
    \setlength{\abovecaptionskip}{6pt}
    \centering
    \includegraphics[width=1\linewidth]{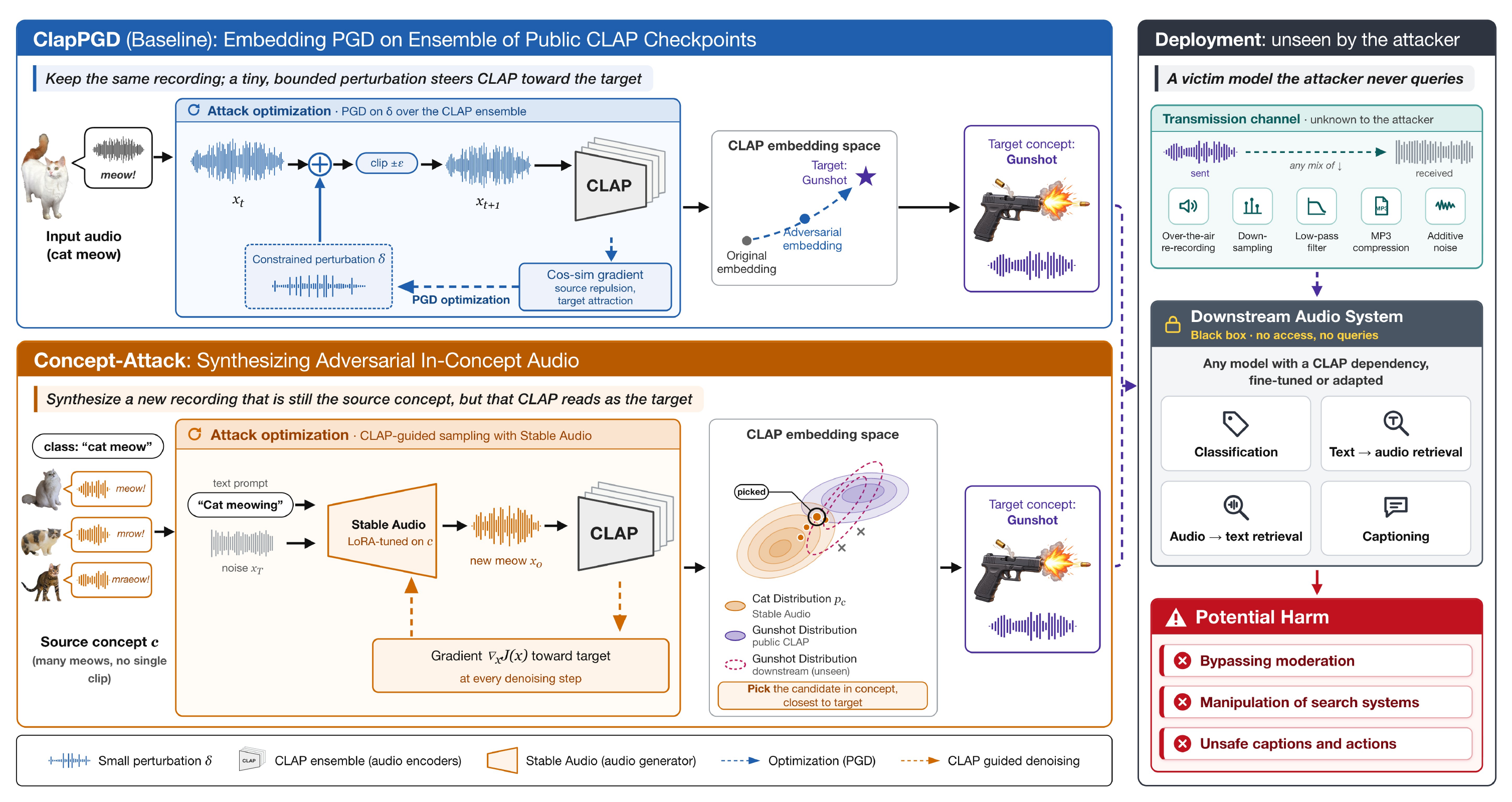}
    \caption{Illustration: how \textsc{clapPGD} (\emph{top}) \& \textsc{ConceptAttack} (\emph{bottom}) attacks corrupt downstream systems.}
    \label{fig:attack_diagram}
    % Tighten the transition to the text for this figure only.
    \vspace{-14pt}
\end{figure*}

Both attacks face the same unusual transfer problem: every attack is optimized only against public CLAP checkpoints and must then transfer to an unseen downstream system. The attacker constructs no surrogate for the target and receives no feedback from it. \S~\ref{sec:threat-model} states the attacker's full knowledge and access constraints.

\subsection{Perturbation: \textsc{clapPGD}}
In the constrained case, the adversary must preserve a particular source waveform $x_s$. \textsc{clapPGD} constructs
% Keep the short waveform equation close to its explanatory sentence.
\begingroup
\setlength{\abovedisplayskip}{3pt}
\setlength{\belowdisplayskip}{3pt}
\setlength{\abovedisplayshortskip}{0pt}
\setlength{\belowdisplayshortskip}{1.5pt}
\[
x_{\mathrm{adv}} = x_s + \delta
\]
and constrains $\delta$ to a small perturbation region.
\par\endgroup

\textsc{clapPGD} optimizes $\delta$ with Projected Gradient Descent (PGD). PGD ordinarily attacks the model whose output is being manipulated, or a surrogate chosen for transfer to that model. In contrast, \textsc{clapPGD} applies PGD only to public CLAP checkpoints. The attack is optimized jointly over multiple CLAP checkpoints and evaluated only afterward on unseen downstream systems.

To construct targeted adversarial examples, we jointly repel the adversarial representation from the source representation and attract it toward an attacker-chosen target representation. \textsc{clapPGD} uses audio centroids for both source repulsion and target attraction.

\medskip\noindent\textbf{Optimizing for \emph{query-free} downstream transfer.}
It is well-known that perturbations optimized against a single checkpoint are prone to overfitting, and this transfer problem is particularly severe in audio (e.g \cite{abdullah_demystifying_nodate,hussain_soundbreak_2026}). \textsc{clapPGD} thus optimizes jointly over an ensemble of public CLAP checkpoints \cite{elizalde_natural_2024,elizalde_clap_2023,wu_large-scale_2024}, as prior work shows that ensembles improve audio adversarial transfer \cite{guo_towards_2023}. We apply dropout during optimization, exposing the attack to local variations of each checkpoint. Both techniques use only the public CLAP models; they (atypically) require no surrogate model for the full downstream target, or any queries or information from it.

\subsection{Synthesis: \textsc{ConceptAttack}}
\label{sec:concept_methodology}
We instantiate \emph{Synthesis} with an attack designed for strong downstream transfer to CLAP-derived systems.

Inspired by the recent concept-based image attack framework of Zhang et al. \cite{zhang_concept-based_2026}, we design the first concept attacks for audio. Specifically, \textsc{ConceptAttack} first learns a generative distribution $p_c$ over realizations of a source concept $c$. It then guides samples from $p_c$ against the public CLAP ensemble toward an attacker-chosen target concept. Thus, a targeted \texttt{meow2glock} attack (i.e., `meow' $\rightarrow$ `glock') searches over newly generated realizations of `meow' for audio that CLAP encodes toward `glock'. We use this larger synthesis search space to improve transfer across more distant CLAP-derived systems, where bounded waveform perturbations may overfit to public CLAPs. \textsc{ConceptAttack} introduces a novel independent audio model evaluation during attack construction that preserves the source concept while exploiting this large search space.

\medskip\noindent\textbf{Source-concept distribution.}
\textsc{ConceptAttack} constructs a distribution $p_c$ for each source concept $c$ by LoRA fine-tuning Stable Audio Open 1.0~\cite{evans_stable_2025}. For ESC-50 and UrbanSound8K, each class defines a concept, and we fine-tune on the training examples from that class.

The AudioCaps and Clotho datasets contain individual recordings rather than class-level concepts. Following Zhang et al. \cite{zhang_concept-based_2026}, we expand each selected recording into a small concept set before fine-tuning. Specifically, we randomly sample $100$ recordings. For each source recording, we paraphrase its caption into $50$ variations with GPT-4o-mini and use those prompts, together with the original audio, to generate corresponding realizations with Stable Audio 3 Medium~\cite{evans_stable_2026}. These generated recordings form the concept set used to train $p_c$. For evaluation, we assign each of the $100$ source recordings ten distinct randomly sampled target recordings from the full dataset, excluding the source itself, yielding $1{,}000$ source--target trials per dataset. We generate $M$ candidates for each trial and select one for evaluation. Retrieval is evaluated against the full official gallery.

\medskip\noindent\textbf{CLAP-guided attack generation objective.}
The \textsc{ConceptAttack} objective is defined entirely over the public CLAP ensemble. At each denoising step, \textsc{ConceptAttack} guides the sampled audio $x$ using a gradient $\nabla_x J(x)$ of an objective $J$ evaluated over that ensemble. Let $E_m^{\mathrm{audio}}$ and $E_m^{\mathrm{text}}$ denote audio and text encoders of public CLAP model $m$, and let $w_m$ denote its ensemble weight.

We consider two attacker-selected objectives over CLAP itself. One targets CLAP's zero-shot score for an attacker-chosen class prompt; the other targets CLAP audio--text similarity to attacker-chosen text. We evaluate the former on ESC-50 and UrbanSound8K and the latter on AudioCaps and Clotho. Neither choice requires knowledge of the downstream task or system.

For the zero-shot objective, the targeted attack specifies a target \emph{class} prompt and maximizes its CLAP score:
\begin{equation}
J_{\mathrm{ZSC}}(x)
=
\sum_m w_m
\log
\frac{
    \exp\!\left(\tau\,s_{m,t}(x)\right)
}{
    \sum_c \exp\!\left(\tau\,s_{m,c}(x)\right)
},
\end{equation}
where $s_{m,c}(x)$ is the similarity between audio $x$ and class prompt $c$ under CLAP model $m$, and $\tau$ is a temperature.

For the audio--text objective, the targeted attack specifies target \emph{text} and maximizes audio--text similarity to its CLAP embedding:
\begin{equation}
J_{\mathrm{retrieval}}(x)
=
\sum_m w_m
\cos\!\left(
    E_m^{\mathrm{audio}}(x),
    E_m^{\mathrm{text}}(c_t)
\right),
\end{equation}
where $c_t$ is the attacker-chosen target caption.

Both objectives require only public CLAP. They make no assumption about how the downstream system will consume the resulting representation.

\subsubsection{Automated concept preservation}
\label{sec:judges}

Strong CLAP guidance can yield candidates that deviate from the source-concept distribution. \textsc{ConceptAttack} rejects candidates whose source-concept confidence under independent judges falls below a fixed threshold.

The original concept-based attack formulation in vision verifies concept preservation after generation using human evaluation \cite{zhang_concept-based_2026}. \textsc{ConceptAttack} automates this constraint during attack construction. Specifically, we use two independent audio representation models, OpenBEATS-Large \cite{bharadwaj_openbeats_2025} and Audio-MAE++ Large \cite{yadav_audiomae_2025}, to score each generated candidate for consistency with the source concept. Their architectures and training objectives differ substantially from CLAP, reducing the likelihood that concept preservation is evaluated via the same representation-level failure mode that attacks target. Table~\ref{tab:judges_performance} reports linear-probe accuracy on ESC-50 and UrbanSound8K and cosine-similarity AUROC on AudioCaps and Clotho.

\begin{table}[h]
\centering
\caption{Performance of concept preservation judges, measured via mean Top-1 linear-probe accuracy with frozen embeddings or cosine-similarity AUROC.
%Values report mean linear-probe accuracy with frozen embeddings on ESC-50 and US8K and cosine-similarity AUROC on AudioCaps and Clotho.
}
\label{tab:judges_performance}
\begin{tabular}{lcccc}
\toprule
& \multicolumn{2}{c}{\textbf{Top-1 Acc. (\%)}}
& \multicolumn{2}{c}{\textbf{AUROC (\%)}} \\
\cmidrule(lr){2-3}\cmidrule(lr){4-5}
\textbf{Judge} & ESC-50 & US8K & AudioCaps & Clotho \\
\midrule
OpenBEATs Large  & 89.0 & 84.7 & 98.3 & 99.1 \\
AudioMAE++ Large & 90.2 & 84.6 & 98.9 & 98.0 \\
\bottomrule
\end{tabular}
\end{table}

\medskip\noindent\textbf{Candidate selection.}
Following Zhang et al. \cite{zhang_concept-based_2026}, \textsc{ConceptAttack} generates $M$ guided candidates from $p_c$ for each source--target pair; we use $M=10$. Each candidate must satisfy two conditions: strong adversarial alignment under the public CLAP ensemble and preservation of the source concept under the independent judges. Let $A(x)$ denote a candidate's adversarial score on the CLAP ensemble, and let $J_{\text{judge}}(x)$ denote the maximum source-concept confidence assigned by either preservation judge. Candidates below the preservation threshold are discarded; the strongest remaining attack is returned:
\begin{equation}
x^\star = \arg\max_{x \in \mathcal{V}} A(x), \qquad
\mathcal{V} = \{x_i : J_{\text{judge}}(x_i) \geq 0.9\}.
\end{equation}

Here, the preservation threshold requires at least one judge to classify the candidate as the source concept with confidence $\geq 90\%$. Thus, source-concept confidence constrains attack construction directly.
For AudioCaps and Clotho, confidence is candidate--source embedding cosine similarity with $\tau=0.90$; Table~\ref{tab:judges_performance} reports threshold-independent AUROC.

\subsection{Ensemble of Models}
We generate all the attacks using an equally weighted ensemble of four public CLAP checkpoints:
 \textbf{MS-CLAP-22}~\cite{elizalde_clap_2023},  \textbf{MS-CLAP-23}~\cite{elizalde_natural_2024}, \textbf{LAION-CLAP} (\texttt{630k-audioset-best.pt})~\cite{wu_large-scale_2024}, and \textbf{LAION-CLAP} (\texttt{630k-audioset-fusion-best.pt})~\cite{wu_large-scale_2024}. Each surrogate contributes equally to the attack objective, as the attacker assumes that the downstream system relies on CLAP, but does not know which public checkpoint is ultimately used (or which serves as the initialization for a fine-tuned descendant, etc.).
    % \noindent\textbf{\textsc{clapPGD}.} A generic contrastive objective that simultaneously attracts the adversarial example toward a target embedding while repelling it from the original embedding.
    % \item \textbf{CrossFire.} We adapt the CrossFire attack from \cite{dou_adversarial_2024} to the CLAP embedding space. Unlike Adversarial Illusions, CrossFire first transforms the target into the same modality as the source before optimizing the adversarial example toward the transformed target, reducing cross-modal alignment errors during optimization.

% \input{Sections/ExperimentsSetup-AB4}
\section{Experimental Setup}
\label{sec:experimental-setup}

Our evaluation investigates whether adversarial audio constructed
against public CLAP checkpoints transfers to previously unseen
downstream systems, and whether concept-based attacks exhibit
stronger downstream transfer than conventional waveform
perturbation attacks.

We evaluate transfer across systems with different forms of CLAP
dependency, spanning frozen encoder reuse, fine-tuned CLAP-pretrained
encoders, and independently trained CLAP-family encoders.
Our evaluation encompasses four public CLAP checkpoints, five
downstream systems, four datasets, and four evaluation tasks.
No downstream evaluation model is queried during attack construction.

\Needspace{18\baselineskip}
\subsection{Downstream Models and CLAP Dependencies}
\label{sec:downstream-models}

% Keep the wrapped table close to its text without affecting other floats.
\begingroup
\setlength{\intextsep}{4pt}
\begin{wraptable}{r}{0.58\linewidth}
    \centering
    \setlength{\abovecaptionskip}{0pt}
    \caption{CLAP dependencies of evaluated downstream systems.}
    \label{tab:model-dependencies}
    \footnotesize
    \setlength{\tabcolsep}{3pt}
    \renewcommand{\arraystretch}{1.25}
    \begin{tabularx}{\linewidth}{@{}ll
        >{\raggedright\arraybackslash}X@{}}
        \toprule
        \textbf{Model} & \textbf{Audio encoder} & \textbf{CLAP dependency} \\
        \midrule
        EnCLAP  & LAION-CLAP Fusion
                & Frozen; additional audio-codec features \\
        \addlinespace[4pt]
        ClapCap & MS-CLAP-23
                & Frozen; learned mapper \\
        \addlinespace[4pt]
        AF1     & MS-CLAP-23 via ClapCap
                & Frozen; integrated into audio-language model \\
        \midrule
        Pengi   & CLAP-pretrained HTSAT
                & Fine-tuned \\
        \midrule
        AF2     & AF-CLAP
                & Independently trained \\
        \bottomrule
    \end{tabularx}
\end{wraptable}

We distinguish between attack success on the four public CLAP
checkpoints used during attack construction and adversarial
transfer to five previously unseen downstream systems.

As summarized in Table~\ref{tab:model-dependencies}, we select
downstream systems with varying forms of CLAP dependency.
EnCLAP \cite{kim_enclap_2024}, ClapCap
\cite{elizalde_natural_2024}, and Audio Flamingo 1 (AF1)
\cite{kong_audio_2024} incorporate frozen CLAP representations
into larger architectures, while Pengi
\cite{deshmukh_pengi_2024} fine-tunes its CLAP-pretrained
audio encoder.

Audio Flamingo 2 (AF2) \cite{ghosh_audio_2025} represents a
particularly challenging transfer setting, as its independently
trained AF-CLAP encoder shares no attacked public checkpoint
parameters.

This selection allows us to evaluate whether adversarial
transfer survives both the integration of CLAP into larger
architectures and the adaptation or independent training of
CLAP-family encoders. For retrieval, we additionally evaluate
AF-CLAP directly in its audio--text embedding space.
\par\endgroup

\subsection{Datasets and Downstream Tasks}

We evaluate adversarial transfer across four widely used audio datasets and three downstream task families: classification, cross-modal retrieval, and captioning. For zero-shot audio classification, we use ESC-50 \cite{piczak_esc_2015} and UrbanSound8K \cite{noauthor_dataset_nodate}, which contain class-labeled environmental audio. For cross-modal retrieval and automated audio captioning, we use AudioCaps \cite{kim_audiocaps_2019} and Clotho \cite{drossos_clotho_2019}, which pair audio recordings with natural-language descriptions.

We consider both directions of cross-modal retrieval: audio-to-text (A2T) and text-to-audio (T2A). In A2T retrieval, the adversary controls the audio query, whereas in T2A retrieval, the adversary controls an audio item in the retrieval corpus while the text queries remain unmodified. Across all four tasks, the adversary seeks to redirect the downstream system's interpretation of the source audio toward attacker-chosen target semantics, without knowledge of the downstream task.

For \textsc{ConceptAttack}, the structure of the available data determines the attack objective: we use a class-level zero-shot CLAP objective for class-labeled datasets and a CLAP audio--text similarity objective for captioned datasets. Both objectives are defined independently of the downstream task.
%
% \subsection{Datasets and Downstream Tasks}
%
% We evaluate adversarial transfer across four widely used
% environmental audio datasets and three downstream task families:
% classification, cross-modal retrieval, and captioning.

% \noindent\textbf{Audio classification.}
% We evaluate zero-shot audio classification on ESC-50 and
% UrbanSound8K, measuring whether adversarial audio changes the
% predicted semantic class.

% \noindent\textbf{Cross-modal retrieval.}
% We evaluate both audio-to-text and text-to-audio retrieval on
% AudioCaps and Clotho. In audio-to-text retrieval, the adversary
% controls the audio query. In text-to-audio retrieval, the adversary
% controls an audio item in the retrieval corpus while the text
% queries remain unmodified.

% \noindent\textbf{Audio captioning.}
% We evaluate automated audio captioning on AudioCaps and Clotho,
% measuring whether adversarial audio induces captions that deviate
% from the source semantics or describe attacker-chosen semantics.
%
\subsection{Evaluated Attacks}
We compare our proposed synthesis-based attack,
\textsc{ConceptAttack}, with two perturbation-based baselines.
All three methods construct attacks against the same ensemble
of four public CLAP checkpoints, without access to or feedback
from the downstream evaluation systems.

\noindent\textbf{\textsc{clapPGD}.}
An ensemble waveform PGD baseline that optimizes bounded
perturbations against public CLAP audio representations.
We additionally employ dropout during optimization to improve
transferability.

\noindent\textbf{Adversarial Illusions.}
We adapt the multimodal embedding attack Adversarial Illusions
to the CLAP embedding space. Unlike \textsc{clapPGD}, this
baseline targets text embeddings rather than audio embeddings.

\noindent\textbf{\textsc{ConceptAttack} (Ours).}
Our synthesis-based attack generates adversarial realizations
of the source concept rather than perturbing a fixed source
waveform, as described in Section~\ref{sec:concept_methodology}.

% \medskip\noindent\textbf{Adversarial Illusions Baseline.}
% Although no prior attack targets our threat model, we adapt Adversarial Illusions \cite{zhang_adversarial_2024}, a recent multimodal embedding attack, to the CLAP embedding space, yielding an embedding-space attack baseline constructed on the same public checkpoints.
% % \medskip\noindent\textbf{\textcolor{orange}{MORE BASELINES? WAV source-target 25\% MIX?. This is now REDUNDANT WITH THE MINI SECTION ON ATTACK BASELINES ABOVE - ADAM FIX THIS}}

% All attack methods are constructed without queries to the downstream evaluation systems. Differences in downstream attack success therefore measure transfer from the public CLAP attack surface rather than adaptation to the evaluated target.

\subsection{Evaluation Protocol and Metrics}

All attacks in our evaluation are constructed with targeted
adversarial objectives. We evaluate their effectiveness along
two complementary dimensions: their ability to disrupt the
downstream system's correct interpretation of the source audio
(\emph{untargeted metrics}) and their ability to redirect its
interpretation toward attacker-chosen target semantics
(\emph{targeted metrics}).
Both dimensions are evaluated on the same set of adversarial
examples.

% \noindent\textbf{Clean controls.}
% For perturbation-based attacks, the clean control is the original,
% unperturbed waveform.
% For classification with \textsc{ConceptAttack}, the clean control
% is audio synthesized using the same random seed but without
% adversarial CLAP guidance.
% This accounts for classification errors arising from synthesis
% itself rather than adversarial guidance.

\noindent\textbf{Classification.}
We report top-1 and top-5 accuracy with respect to the source class, as well as untargeted and targeted attack success rates (ASR). Untargeted ASR measures the fraction of clean-correct examples that are misclassified after the attack. For perturbation-based attacks, the clean control is the original waveform; for \textsc{ConceptAttack}, it is audio synthesized using the same random seed without adversarial CLAP guidance. Targeted ASR measures the fraction of all adversarial examples classified as the attacker-chosen target class.

\noindent\textbf{Retrieval.}
For both audio-to-text and text-to-audio retrieval, we report untargeted and targeted ASR at retrieval cutoffs $K \in \{1, 5, 10\}$. Untargeted ASR@$K$ measures the fraction of attacks that remove the correct source item from the top-$K$ results, conditioned on successful clean retrieval at the same cutoff. Targeted ASR@$K$ measures the fraction of attacks for which the designated target item appears among the top-$K$ results. All retrieval experiments use the full evaluation gallery.

\noindent\textbf{Captioning.}
Unlike classification and retrieval, captioning produces open-ended natural language outputs, requiring metrics that capture the degree of semantic redirection. To measure untargeted effectiveness, we evaluate agreement with the source reference captions using SPIDEr-FL, CIDEr-D, SPICE, and FENSE, where lower scores indicate greater disruption of the original interpretation. To measure targeted effectiveness, we evaluate FENSE against the attacker-chosen target caption, where higher scores indicate stronger alignment with the target semantics. FENSE measures semantic agreement between captions while penalizing fluency errors, and SPIDEr-FL incorporates a fluency penalty into SPIDEr \cite{zhou_can_2022}.

\section{Results}
%
%
% Requires: \usepackage{booktabs,tabularx,array,xcolor}
% \definecolor{ConceptBenign}{HTML}{1B5E20}
% \colorlet{ConceptHarm}{red}
%  NOTE I MOVED THE COLOR DEFS TO THE PREAMBLE AS THEY ARE USED IN MULTPLE TABLES NOW incl intro

\newcommand{\ConceptExample}[3]{%
  \begin{minipage}[t]{\linewidth}
    \textit{Clean:} #1\par
    \textit{Target:} #2\par
    \textbf{Adversarial:} #3
  \end{minipage}%
}

\begin{table*}[t]
\centering
\caption{Selected qualitative examples of \textsc{ConceptAttack} captioning on AudioCaps
and Clotho.}
\label{tab:concept-attack-caption-examples}
\footnotesize
\setlength{\tabcolsep}{4pt}
\renewcommand{\arraystretch}{1.00}
\begin{tabularx}{\textwidth}{@{}>{\raggedright\arraybackslash}p{1.8cm}
  >{\raggedright\arraybackslash}X
  >{\raggedright\arraybackslash}X@{}}
\toprule
\textbf{Model} &
\multicolumn{1}{c}{\textbf{\textcolor{ConceptBenign}{Benign}
  $\boldsymbol{\rightarrow}$ \textcolor{ConceptHarm}{Harm}}} &
\multicolumn{1}{c}{\textbf{\textcolor{ConceptHarm}{Harm}
  $\boldsymbol{\rightarrow}$ \textcolor{ConceptBenign}{Benign}}} \\
\cmidrule(r){1-1}\cmidrule(lr){2-2}\cmidrule(l){3-3}

\DownstreamModelCell{AF1}{Frozen} &
\ConceptExample{A train horn blows, and a train passes by.}
  {Loading and firing of a gun, followed by footsteps and speech.}
  {A person is talking and \textbf{\textcolor{ConceptHarm}{a gun is fired}},
   then a person is talking and \textbf{\textcolor{ConceptHarm}{a gun is fired again}}.}
&
\ConceptExample{A \textbf{\textcolor{ConceptHarm}{firework explodes}} in the sky and then fades away.}
  {Cars pass on a road while a dog barks.}
  {\textbf{\textcolor{ConceptBenign}{A dog is barking and a car is driving by.}}}
\\[2pt]
\midrule

\DownstreamModelCell{ClapCap}{Frozen} &
\ConceptExample{A person uses a saw to cut a piece of wood.}
  {A siren rings as traffic passes.}
  {\textbf{\textcolor{ConceptHarm}{A siren blares and a siren goes off.}}}
&
\ConceptExample{\textbf{\textcolor{ConceptHarm}{Fireworks are going off}} in the distance.}
  {Cars pass on a road while a dog barks.}
  {\textbf{\textcolor{ConceptBenign}{A dog barks and then a loud dog barks.}}}
\\[2pt]
\midrule

\DownstreamModelCell{EnCLAP}{Frozen} &
\ConceptExample{A large volume of water is splashing and gurgling, and people are speaking.}
  {Someone clicks a gun and fires multiple shots.}
  {\textbf{\textcolor{ConceptHarm}{A gun is fired several times.}}}
&
\ConceptExample{\textbf{\textcolor{ConceptHarm}{A child cries and squeals}} as birds chirp.}
  {Typing on a keyboard.}
  {\textbf{\textcolor{ConceptBenign}{Typing on a computer keyboard.}}}
\\[2pt]
\midrule

\DownstreamModelCell{Pengi}{Fine-tuned} &
\ConceptExample{A dog is barking and a dog is crying.}
  {Rapid gunfire with a distant voice.}
  {\textbf{\textcolor{ConceptHarm}{A gun is being fired at a target.}}}
&
\ConceptExample{\textbf{\textcolor{ConceptHarm}{Fireworks are being fired}} in the distance.}
  {A bird caws while smaller birds chirp in the background.}
  {\textbf{\textcolor{ConceptBenign}{Birds chirping and whistling in the background.}}}
\\[2pt]
\midrule

\DownstreamModelCell{AF2}{Independent} &
\ConceptExample{A person is walking on a wooden floor with a heavy load of wood.}
  {A loud explosion that gradually becomes less intense.}
  {\textbf{\textcolor{ConceptHarm}{A loud explosion}} with deep bass and a long reverb.}
&
\ConceptExample{A person is playing a game of \textbf{\textcolor{ConceptHarm}{fireworks exploding}}.}
  {People walking on a creaky, squeaky wooden floor.}
  {\textbf{\textcolor{ConceptBenign}{A person is walking on a wooden floor with shoes on.}}}

\\[2pt]
\bottomrule
\end{tabularx}
\end{table*}

\label{sec:results}
Across all four tasks, attacks constructed against public CLAP checkpoints transfer to previously unseen systems that depend on them, even peripherally. On the attacked public checkpoints, \textsc{clapPGD} and Adversarial Illusions frequently match or outperform \textsc{ConceptAttack}; downstream, the ordering changes. As the downstream model moves farther from the attacked checkpoints, perturbation transfer falls sharply while \textsc{ConceptAttack} persists. The separation survives even in AF2, whose AF-CLAP encoder shares no attacked public checkpoint parameters.

Table~\ref{tab:captioning} reports results for \emph{Captioning}; we defer full results for \emph{Classification}, \emph{Audio-to-text Retrieval}, and \emph{Text-to-audio Retrieval} to App.~\S\ref{app:fullexperimentalresults}. We highlight three results.

\medskip\noindent\textbf{Captioning attacks can compel the targeted semantics as strongly as clean audio compels the true caption.}
Table~\ref{tab:concept-attack-caption-examples} shows examples of \emph{bypass} and \emph{trigger} attacks generated by \textsc{ConceptAttack}. EnCLAP turns splashing water and speech into ``A gun is fired several times.'' Pengi turns a barking and crying dog into ``A gun is being fired at a target.'' Security-relevant audio can likewise be redirected toward benign captions.

\newcommand{\CaptionCleanLabel}{\textcolor{black!55}{(Clean)}}
\newcommand{\CaptionCleanValue}[1]{\textcolor{black!55}{(#1)}}

% Keep the table together at the top of a text page so the remaining
% space below it can hold the surrounding discussion.
\begin{table*}[!t]
\centering
\caption{Downstream evaluation on automated audio captioning. \emph{Clean} reports unattacked performance.}

\small
\setlength{\tabcolsep}{2.8pt}
\renewcommand{\arraystretch}{1.02}
\begin{tabular*}{\textwidth}{@{\extracolsep{\fill}}lllccccc@{}}
\toprule
& & &
\multicolumn{4}{c}{\textbf{Untargeted}} &
\multicolumn{1}{c}{\textbf{Targeted}} \\
\cmidrule(lr){4-7}
\cmidrule(lr){8-8}
\textbf{Dataset} &
\textbf{Model} &
\textbf{Attack} &
\textbf{SPIDEr-FL $\downarrow$} &
\textbf{CIDEr-D $\downarrow$} &
\textbf{SPICE $\downarrow$} &
\textbf{FENSE $\downarrow$} &
\textbf{FENSE $\uparrow$} \\
\midrule 

\multirow{20}{*}{\rotatebox[origin=c]{90}{\textbf{AudioCaps}}}
& \multirow{4}{*}{\DownstreamModelCell{EnCLAP}{Frozen}}
& \CaptionCleanLabel
& \CaptionCleanValue{.484}
& \CaptionCleanValue{.762}
& \CaptionCleanValue{.217}
& \CaptionCleanValue{.650}
& \CaptionCleanValue{.220} \\
& & Adv.\ Illusions
& \textbf{.048}
& \textbf{.062}
& \textbf{.034}
& .219
& \textbf{.651} \\
& & \textsc{clapPGD}
& .051
& .068
& \textbf{.034}
& \textbf{.217}
& .610 \\
& & \textsc{ConceptAttack}
& .055
& .074
& .038
& .232
& .620 \\
\cmidrule(lr){2-8}\addlinespace[0.1em]

& \multirow{4}{*}{\DownstreamModelCell{ClapCap}{Frozen}}
& \CaptionCleanLabel
& \CaptionCleanValue{.286}
& \CaptionCleanValue{.431}
& \CaptionCleanValue{.172}
& \CaptionCleanValue{.519}
& \CaptionCleanValue{.189} \\
& & Adv.\ Illusions
& .252
& .388
& .148
& .479
& .199 \\
& & \textsc{clapPGD}
& .227
& .341
& .136
& .456
& .214 \\
& & \textsc{ConceptAttack}
& \textbf{.058}
& \textbf{.077}
& \textbf{.045}
& \textbf{.211}
& \textbf{.375} \\
\cmidrule(lr){2-8}\addlinespace[0.1em]

& \multirow{4}{*}{\DownstreamModelCell{AF1}{Frozen}}
& \CaptionCleanLabel
& \CaptionCleanValue{.277}
& \CaptionCleanValue{.415}
& \CaptionCleanValue{.140}
& \CaptionCleanValue{.535}
& \CaptionCleanValue{.185} \\
& & Adv.\ Illusions
& .202
& .290
& .116
& .473
& .231 \\
& & \textsc{clapPGD}
& .146
& .208
& .089
& .392
& .280 \\
& & \textsc{ConceptAttack}
& \textbf{.061}
& \textbf{.073}
& \textbf{.050}
& \textbf{.243}
& \textbf{.471} \\
\cmidrule(lr){2-8}\addlinespace[0.1em]

& \multirow{4}{*}{\DownstreamModelCell{Pengi}{Fine-tuned}}
& \CaptionCleanLabel
& \CaptionCleanValue{.201}
& \CaptionCleanValue{.312}
& \CaptionCleanValue{.112}
& \CaptionCleanValue{.462}
& \CaptionCleanValue{.185} \\
& & Adv.\ Illusions
& .175
& .263
& .107
& .446
& .197 \\
& & \textsc{clapPGD}
& .168
& .251
& .102
& .430
& .211 \\
& & \textsc{ConceptAttack}
& \textbf{.053}
& \textbf{.067}
& \textbf{.043}
& \textbf{.210}
& \textbf{.348} \\
\cmidrule(lr){2-8}\addlinespace[0.1em]

& \multirow{4}{*}{\DownstreamModelCell{AF2}{Independent}}
& \CaptionCleanLabel
& \CaptionCleanValue{.289}
& \CaptionCleanValue{.427}
& \CaptionCleanValue{.151}
& \CaptionCleanValue{.580}
& \CaptionCleanValue{.194} \\
& & Adv.\ Illusions
& .263
& .383
& .150
& .552
& .198 \\
& & \textsc{clapPGD}
& .255
& .369
& .146
& .538
& .205 \\
& & \textsc{ConceptAttack}
& \textbf{.062}
& \textbf{.077}
& \textbf{.048}
& \textbf{.201}
& \textbf{.339} \\

\addlinespace[0.10em]
\midrule
\addlinespace[0.05em]

\multirow{20}{*}{\rotatebox[origin=c]{90}{\textbf{Clotho}}}
& \multirow{4}{*}{\DownstreamModelCell{EnCLAP}{Frozen}}
& \CaptionCleanLabel
& \CaptionCleanValue{.148}
& \CaptionCleanValue{.231}
& \CaptionCleanValue{.069}
& \CaptionCleanValue{.497}
& \CaptionCleanValue{.171} \\
& & Adv.\ Illusions
& .020
& .025
& .016
& .195
& .398 \\
& & \textsc{clapPGD}
& \textbf{.018}
& \textbf{.022}
& \textbf{.014}
& \textbf{.190}
& .431 \\
& & \textsc{ConceptAttack}
& .027
& .035
& .021
& .231
& \textbf{.472} \\
\cmidrule(lr){2-8}\addlinespace[0.1em]

& \multirow{4}{*}{\DownstreamModelCell{ClapCap}{Frozen}}
& \CaptionCleanLabel
& \CaptionCleanValue{.504}
& \CaptionCleanValue{.826}
& \CaptionCleanValue{.192}
& \CaptionCleanValue{.610}
& \CaptionCleanValue{.144} \\
& & Adv.\ Illusions
& .395
& .643
& .163
& .547
& .156 \\
& & \textsc{clapPGD}
& .357
& .587
& .150
& .512
& .180 \\
& & \textsc{ConceptAttack}
& \textbf{.072}
& \textbf{.099}
& \textbf{.054}
& \textbf{.221}
& \textbf{.371} \\
\cmidrule(lr){2-8}\addlinespace[0.1em]

& \multirow{4}{*}{\DownstreamModelCell{AF1}{Frozen}}
& \CaptionCleanLabel
& \CaptionCleanValue{.414}
& \CaptionCleanValue{.672}
& \CaptionCleanValue{.157}
& \CaptionCleanValue{.580}
& \CaptionCleanValue{.150} \\
& & Adv.\ Illusions
& .305
& .492
& .128
& .501
& .187 \\
& & \textsc{clapPGD}
& .243
& .384
& .106
& .438
& .252 \\
& & \textsc{ConceptAttack}
& \textbf{.069}
& \textbf{.092}
& \textbf{.048}
& \textbf{.232}
& \textbf{.440} \\
\cmidrule(lr){2-8}\addlinespace[0.1em]

& \multirow{4}{*}{\DownstreamModelCell{Pengi}{Fine-tuned}}
& \CaptionCleanLabel
& \CaptionCleanValue{.336}
& \CaptionCleanValue{.543}
& \CaptionCleanValue{.140}
& \CaptionCleanValue{.527}
& \CaptionCleanValue{.139} \\
& & Adv.\ Illusions
& .279
& .449
& .119
& .481
& .159 \\
& & \textsc{clapPGD}
& .247
& .394
& .108
& .457
& .189 \\
& & \textsc{ConceptAttack}
& \textbf{.065}
& \textbf{.087}
& \textbf{.047}
& \textbf{.208}
& \textbf{.372} \\
\cmidrule(lr){2-8}\addlinespace[0.1em]

& \multirow{4}{*}{\DownstreamModelCell{AF2}{Independent}}
& \CaptionCleanLabel
& \CaptionCleanValue{.194}
& \CaptionCleanValue{.315}
& \CaptionCleanValue{.113}
& \CaptionCleanValue{.440}
& \CaptionCleanValue{.130} \\
& & Adv.\ Illusions
& .194
& .308
& .104
& .436
& .147 \\
& & \textsc{clapPGD}
& .195
& .310
& .103
& .428
& .153 \\
& & \textsc{ConceptAttack}
& \textbf{.062}
& \textbf{.079}
& \textbf{.049}
& \textbf{.209}
& \textbf{.293} \\

\bottomrule
\end{tabular*}
\label{tab:captioning}
\end{table*}

In mean attack success on AudioCaps, \textsc{ConceptAttack} reaches target FENSE of $.375$, $.471$, and $.348$ on ClapCap, AF1, and Pengi, compared with $.214$, $.280$, and $.211$ for \textsc{clapPGD}. On Clotho, the corresponding values are $.371$, $.440$, and $.372$, compared with $.180$, $.252$, and $.189$. EnCLAP provides the clearest scale: Adversarial Illusions obtains $.651$ FENSE against the attacker-selected target caption, while clean EnCLAP captions obtain $.650$ against their true references; \textsc{clapPGD} and \textsc{ConceptAttack} reach $.610$ and $.620$ against their targets.

\medskip\noindent\textbf{Attacks survive downstream reuse and adaptation.}
On ESC-50, \textsc{ConceptAttack} flips $86.2\%$ of clean-correct predictions on fine-tuned Pengi and $88.0\%$ on frozen AF1, compared with $49.8\%$ and $71.2\%$ for \textsc{clapPGD}. On UrbanSound8K, \textsc{ConceptAttack} flips $71.9\%$ and $73.3\%$, compared with $50.5\%$ and $54.9\%$ for \textsc{clapPGD}. Similarly, for targeted classification \textsc{ConceptAttack} obtains $31.8\%/45.4\%$ on ESC-50 and $27.8\%/35.6\%$ on UrbanSound8K for Pengi/AF1, versus $14.3\%/26.5\%$ and $19.9\%/28.8\%$ for \textsc{clapPGD}.

A natural concern is that the attacks overfit to public CLAP parameters and may disappear after fine-tuning. They do not. We fine-tune five independently seeded LAION-CLAP models for $20$ epochs on AudioCaps and re-evaluate the same attacks throughout fine-tuning. Targeted ASR declines only slightly across the full fine-tuning trajectory (App.~\S\ref{app:finetuningnotadefense}). 

\medskip\noindent\textbf{\textsc{ConceptAttack} crosses the no-shared-checkpoint boundary.}
AF2 provides the strongest challenge in our evaluation because it is independent of public CLAP checkpoint parameters. Nevertheless, \textsc{ConceptAttack} flips $69.2\%$ of clean-correct ESC-50 predictions and $60.9\%$ of UrbanSound8K predictions, compared with $18.3\%$ and $19.1\%$ for \textsc{clapPGD}. For targeted, the divide is slimmer on ESC-50 with ($8.2\%$ vs.\ $3.3\%$), yet on UrbanSound8K ($31.1\%$ vs.\ $13.1\%$).

Retrieval produces the clearest illustration of the \textsc{ConceptAttack} transfer advantage. For audio-to-text retrieval on AudioCaps, \textsc{ConceptAttack} removes the source from Audio Flamingo 2's top-$10$ results in $91.0\%$ of queries and reaches $30.4\%$ targeted ASR@10; \textsc{clapPGD} reaches only $7.7\%$ and $0.6\%$. On Clotho, \textsc{ConceptAttack} reaches $98.5\%$ untargeted and $14.6\%$ targeted ASR@10, while neither perturbation baseline obtains a targeted top-$10$ success. For text-to-audio retrieval on AudioCaps, attacker-chosen audio enters AF-CLAP's top-$10$ results in $52.0\%$ of targeted attacks under \textsc{ConceptAttack}, compared with $6.1\%$ for \textsc{clapPGD} and $3.8\%$ for Adversarial Illusions. On Clotho, \textsc{ConceptAttack} reaches $100\%$ untargeted ASR@10 and $8.9\%$ targeted ASR@10, compared with at most $10.1\%$ and $0.5\%$, respectively, for the perturbation baselines.

\textsc{ConceptAttack} also survives into AF2's generated captions: target FENSE reaches $.339$ on AudioCaps and $.293$ on Clotho, compared with $.205/.153$ for \textsc{clapPGD} and $.198/.147$ for Adversarial Illusions. The no-shared-checkpoint boundary does not eliminate downstream transfer. It changes which attacks survive it.

\section{Attacks Survive Digital Signal Degradation and Physical Playback Outdoors on Cheap Consumer Hardware (iPhone)}
\label{sec:signal-degradation}
We show that our attacks survive aggressive signal degradation. We consider two varieties: (1) digital processing (MP3 compression, resampling, low-pass filtering, and white noise); and (2) playback-and-re-recording via a cheap consumer speaker outdoors in the noisy Hollywood Vinyl (recording) District. We emphasize that the attacks are constructed with no knowledge of this downstream signal path. Under these suboptimal conditions, both \textsc{clapPGD} and \textsc{ConceptAttack} survive. As the signal is increasingly degraded, however, \textsc{ConceptAttack} survives substantially better.

We focus these degradation studies on classification, the downstream task most naturally exposed to live environmental capture: deployed systems may need to determine whether a gunshot, alarm, or other environmental sound occurred, including accessibility systems that continuously recognize environmental sounds for deaf and hard-of-hearing users \cite{apple_recognize_nodate}.

\begin{figure*}[t]
    \centering
    \includegraphics[width=1.0\textwidth]{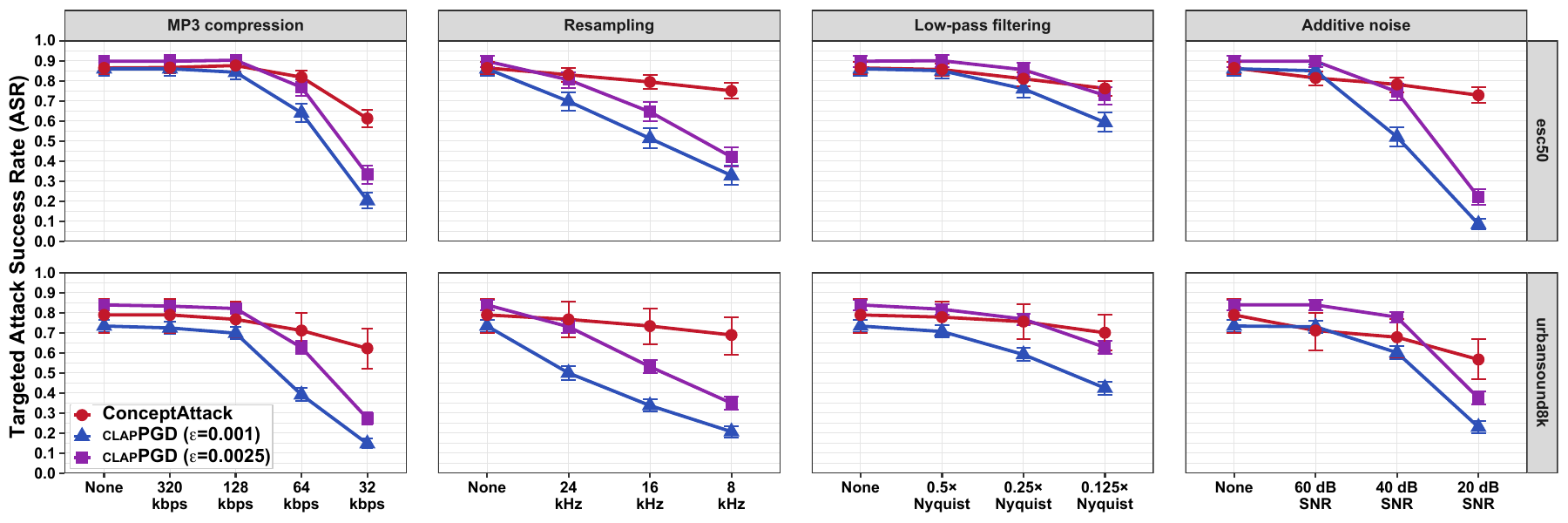}
    \caption{Targeted attack success on LAION-CLAP under digital transformations at varying strengths. \textsc{ConceptAttack}'s wider error bars on UrbanSound8K reflect this dataset's fewer classes (\textsc{ConceptAttack}'s attacks are class-sampled).}
    \label{fig:signal-processing}
\end{figure*}

\medskip\noindent\textbf{\textsc{ConceptAttack} survives aggressive digital processing.}
Fig.~\ref{fig:signal-processing} applies four common transformations at progressively stronger settings: MP3 compression down to $32$ kbps, round-trip resampling down to $8$ kHz, low-pass filtering down to $0.125\times$ Nyquist, and additive white noise down to $20$ dB SNR. Across ESC-50 and UrbanSound8K, \textsc{ConceptAttack} degrades substantially more slowly than \textsc{clapPGD}. Even at the strongest setting of every transformation, \textsc{ConceptAttack} retains substantial targeted ASR; for example, after resampling to $8$ kHz it retains roughly $74\%$ targeted ASR on both datasets, while the perturbation attacks deteriorate  further. The same broad relationship appears under aggressive MP3 compression, additive noise, and low-pass filtering.

% \textcolor{orange}{Extra writeup can be dropped maybe redundant}{\emph{Attacked Ensemble} denotes the public CLAP checkpoints used as direct attack targets; \emph{Downstream} denotes CLAP-family downstream systems evaluated under frozen, fine-tuned, and independently developed settings. For Audio Flamingo 2, \textsc{ConceptAttack} retains 14.4--20.0\% targeted ASR across physical recording conditions while \textsc{clapPGD} ranges from 1.25--11.8\%.}

% \noindent\medskip

\begin{figure*}[!t]
    \centering
    \includegraphics[width=\textwidth]{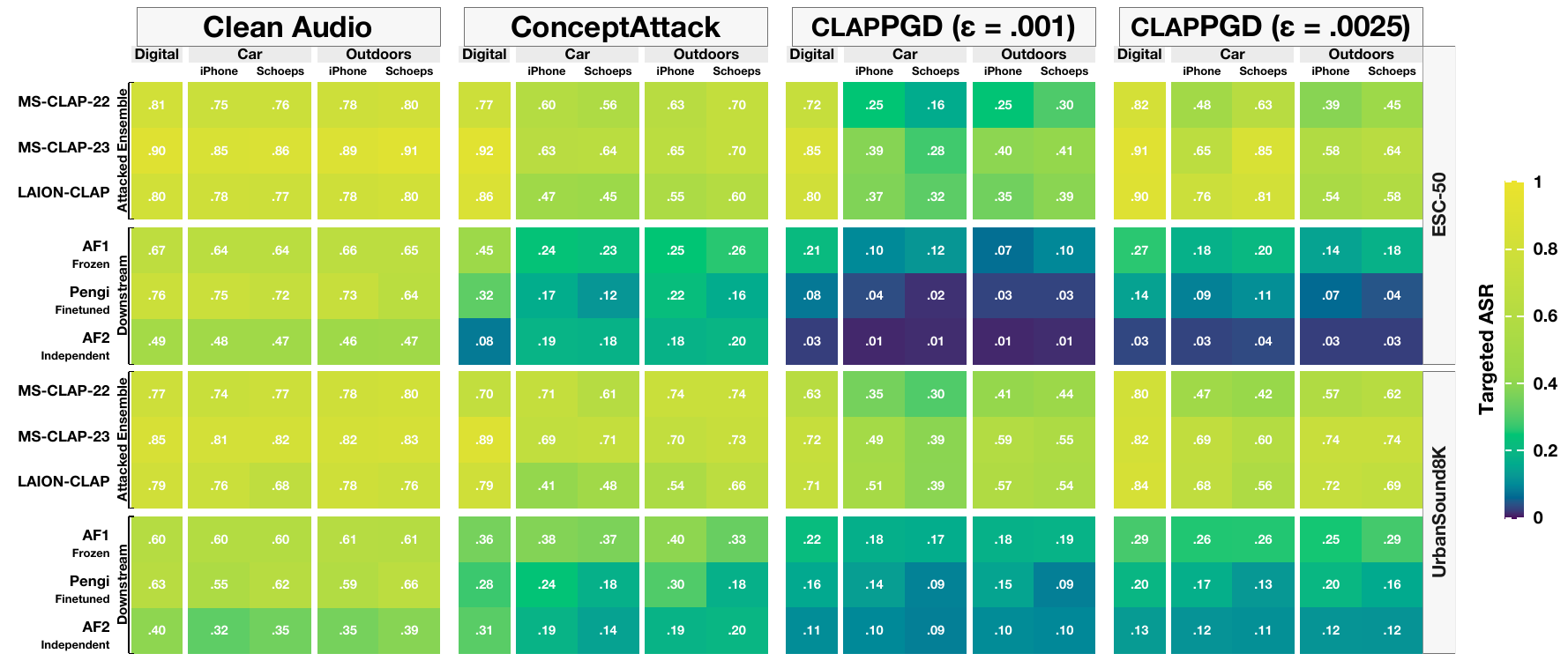}
    \caption{Targeted ASR from digital audio and after over-the-air playback and re-recording. \textsc{clapPGD} obtains stronger digital attacks against public CLAP checkpoints (\emph{panel rows 1 \& 3} \emph{Digital} columns);  \textsc{ConceptAttack} transfers  better to CLAP-family downstream systems and survives better over-the-air (\emph{panel rows 2 \& 4}.  \emph{Clean Audio} reports accuracy.)} 
    \vspace{-1em}
    \label{fig:rerecording-results}
\end{figure*}

\noindent\textbf{The attacks survive physical playback and re-recording.}
\label{sec:liverecording}
We now show that our attacks survive over-the-air playback and re-recording under challenging real-world acoustic conditions and acoustically suboptimal consumer-grade equipment. We consider two environments: (i) a high-ambient-urban-noise \emph{outdoor} location in Hollywood's Vinyl (music recording) District and (ii) the interior of a vehicle. In each environment, we play the attacks using (acoustically suboptimal) consumer-grade playback equipment---a popular 3" consumer-grade studio monitor-speaker. We emphasize the high bar to surviving playback under these acoustically disadvantageous conditions. We re-record the playback using two different capture chains: (1) a professional recording industry-standard Schoeps CMC~1/MK5 condenser microphone, which reflects best-case capture under these suboptimal playback conditions, and (2) the built-in microphone array of the consumer iPhone~16~Pro (with its AI processing disabled). App.~\S\ref{app:liverecording} gives details. We play back and re-record the adversarial examples generated by every attack described in Section~\ref{sec:methods} for both classification datasets: ESC-50 and UrbanSound8K. Figure~\ref{fig:rerecording-results} reports attack performance in both the outdoor and the car environments.

\medskip\textbf{Over-the-air results.}
The attacks survive the physical channel. On the attacked public CLAP checkpoints, the larger-budget \textsc{clapPGD} attack remains strong even after playback and re-recording. Downstream, \textsc{ConceptAttack} again transfers substantially better. %Across the physical recording conditions, \textsc{ConceptAttack} through just an iPhone microphone retains $18$--$19\%$ targeted ASR on AF2, whose AF-CLAP encoder shares no attacked checkpoint parameters; \textsc{clapPGD} ranges from only $1.25$--$11.8\%$.
For reference, AF2's clean digital accuracy is $40$--$49\%$ on these datasets. Even after playback and re-recording with only an iPhone microphone, \textsc{ConceptAttack} still achieves $18$--$19\%$ targeted ASR on this downstream model, despite its AF-CLAP encoder sharing no attacked checkpoint parameters; \textsc{clapPGD} obtains only $1.25$--$11.8\%$.

Most strikingly, the professional capture chain proves unnecessary: attacks constructed only against public CLAP checkpoints survive a consumer speaker, a noisy outdoor acoustic channel, and the iPhone~16~Pro's four-microphone MEMS array---conservatively consumer-priced at \$18.99 for the four-microphone set \cite{ifixit_iphone_2026}---before transferring to a downstream model that shares none of the attacked checkpoint parameters.

% Keep the full-width robustness figures with their discussion.
\FloatBarrier

\section{Related Work}
\label{sec:relatedwork}
Adversarial audio has primarily been studied through waveform perturbations against speech recognition, speaker recognition, and audio classification \cite{qin_imperceptible_2019,abdoli_universal_2020,chen_diffattack_2025}. A persistent challenge is transfer: attacks that succeed against one audio model frequently lose effectiveness against another, motivating ensemble attacks, temporal transformations, and other methods designed specifically to improve black-box transfer \cite{guo_towards_2023,gao_transferable_2024,deng_time_2026}. Recent attacks on multimodal audio systems similarly obtain strong white-box success but substantially weaker transfer across independently trained encoders \cite{hussain_soundbreak_2026}. Our setting differs in the object being attacked. We optimize against public CLAP checkpoints because they are dependencies of the unknown downstream system, then test whether the attack survives through the additional representations, adaptation, and learned components that separate that dependency from the final output.

\medskip\noindent\textbf{Shared representations and downstream transfer.}
Prior work has shown that manipulating a shared representation can propagate across tasks. Adversarial Illusions attacks multimodal embedding spaces and demonstrates effects across retrieval, classification, and generative applications \cite{zhang_adversarial_2024}; CrossFire similarly uses targeted alignment in multimodal representation space \cite{dou_adversarial_2024}. From the defensive direction, RobustCLIP shows that robustness properties of a foundation encoder can propagate into systems that reuse it \cite{schlarmann_robust_2024}. Most closely related to our query-free setting, attacks constructed from public SAM and video foundation models have transferred to unknown fine-tuned downstream models without access to the downstream task or queries \cite{xia_transferable_2024,lu_pretrain_2026}. Our setting pushes this separation further: CLAP is just one frozen feature inside a larger system, a highly fine-tuned ancestor, or even an independently trained CLAP-family encoder with no public/attacked checkpoint parameters.

The difficulty of such transfer is consistent with a broader literature showing that adversarial examples can overfit individual models and that ensembles, checkpoint diversity, and surrogate variation improve transfer \cite{li_learning_2019,gubri_lgv_2022,xie_improving_2019}. More specifically, representation-targeting attacks can fail when source and target representation geometries are misaligned \cite{gupta_understanding_2025}. These results motivate both our ensemble construction and our evaluation across progressively weaker forms of CLAP ancestry.

\medskip\noindent\textbf{Unrestricted and concept-based adversarial examples.}
Conventional adversarial examples search within a constrained neighborhood of an observed input \cite{goodfellow_explaining_2015}. Unrestricted attacks instead synthesize valid adversarial inputs without preserving a particular source example \cite{song_constructing_2018,dai_advdiff_2024}, while related work manipulates semantically meaningful latent attributes rather than unstructured input perturbations \cite{wang_generating_2022}. Most directly, concept-based adversarial attacks model a source concept as a distribution and generate new realizations that preserve that concept while misleading a classifier \cite{zhang_concept-based_2026}. We develop an analogous formulation to the audio modality and use the additional synthesis freedom for a different transfer problem: constructing concept-preserving attacks against public CLAP checkpoints that remain adversarial after passing into unknown downstream systems.

\section{Discussion}
\label{sec:discussion}
\noindent\textbf{Potential defenses.}
Our experiments rule out two natural defenses. Fine-tuning the public checkpoint does not reliably remove transfer, and common signal transformations do not reliably remove \textsc{ConceptAttack}. A more promising intervention may be to harden the public representation itself before release. In vision, RobustCLIP shows that adversarial robustness can propagate from a foundation encoder to downstream systems that reuse it \cite{schlarmann_robust_2024}. Whether robust CLAP training prevents the attacks here remains open. Defenses must also assume an adaptive adversary. To again borrow from vision, LightShed exploits the public availability of popular Glaze and Nightshade attacks to remove their perturbations \cite{shan_glaze_2023,shan_nightshade_2024,foerster_lightshed_2025}. Our setting gives attackers an analogous advantage: attacked CLAP checkpoints are public and can be studied without querying the downstream system.

\medskip\noindent\textbf{Why does \textsc{ConceptAttack} transfer farther?}
We do not yet know. One explanation is the difference in search space: a bounded perturbation searches only a small $\epsilon$-ball around one waveform against particular public checkpoints, whereas \textsc{ConceptAttack} searches over many valid realizations of the source concept. Some realizations may induce the target semantics across multiple related encoders rather than only near one waveform in one checkpoint's geometry. The widening transfer gap as downstream models move farther from the attacked checkpoints, culminating in AF2, is consistent with this explanation. We have not tested the mechanism directly.

\medskip\noindent\textbf{Limitations.}
Our evaluation covers CLAP-family representations; it does not establish transfer to unrelated audio encoders or to speech- and music-specific foundation models. Our automated concept-preservation judge is easiest to define for ESC-50 and UrbanSound8K, where the source concept is a labeled class. While these models provide an automated proxy for semantic consistency, they do not establish that the generated audio preserves the source concept as perceived by human listeners. A systematic human listening study remains future work. AudioCaps and Clotho instead require concepts constructed from captions and generated realizations, which we validate with independent audio models. Finally, synthesis is a more expensive computation than waveform perturbation. \textsc{ConceptAttack}'s greater transfer therefore comes at greater attack cost.

\section{Conclusion}
\label{sec:conclusion}
% In this paper, we have shown that public encoders widely reused across audio AI create an unexpected and broadly applicable attack surface: attacks constructed against the public encoder alone can propagate into downstream systems without knowledge of, access to, or queries to the target system. We also introduce the first concept-based adversarial audio, which generates new audio that remains within a source concept while causing CLAP, and by extension downstream systems, to interpret it as an attacker-chosen target concept. Finally, we show that both PGD and concept-based attacks survive common forms of digital signal degradation and over-the-air playback under challenging acoustic conditions, with concept-based attacks proving particularly robust. As a result, an attacker armed with only a \$250 consumer speaker can manipulate the decisions of an audio AI system they cannot inspect or query and know almost nothing about.
%
In this paper, we have shown that public encoders widely reused across audio AI create an unexpected and broadly applicable attack surface: attacks constructed against the public encoder alone can propagate into downstream systems without knowledge of, access to, or queries to the target system. We also introduce the first concept-based adversarial audio, which preserves a source concept while inducing attacker-chosen semantics downstream. Both attacks survive digital signal degradation, and concept-based attacks survive over-the-air playback into downstream systems that share none of the attacked parameters. As a result, an attacker armed with only a \$250 consumer speaker can manipulate the decisions of an audio AI system they cannot inspect or query and know almost nothing about.

\clearpage

\bibliography{references}
\bibliographystyle{iclr_template/iclr2027_conference}

\clearpage
\appendix
% Keep appendix spacing natural and finish each section's floats before
% starting the next section.
\raggedbottom
\setlength{\parskip}{4pt}
\setlength{\intextsep}{4pt plus 2pt minus 2pt}
\setlength{\textfloatsep}{12pt plus 2pt minus 2pt}
\makeatletter
\setlength{\@fptop}{0pt}
\setlength{\@dblfptop}{0pt}
\makeatother
\section{Experimental Setup Details}
\label{app:experimentalsetup}

\subsection{Downstream Model Details}
\label{app:downstream-model-details}

This section provides additional implementation details for the downstream systems summarized in Table~\ref{tab:downstream-clap-lineage}.

\begin{table}[H]
\centering
\caption{Relationship between downstream evaluation models and public CLAP
checkpoints. ``Exact public ancestor'' denotes the released CLAP checkpoint
whose parameters are reused by the downstream system, where applicable.}
\label{tab:downstream-clap-lineage}
\small
\setlength{\tabcolsep}{2.5pt}
\renewcommand{\arraystretch}{1.12}
\expandafter\def\expandafter\UrlBreaks\expandafter{\UrlBreaks\do\-}
% Reserve real space for every column; the previous fixed widths squeezed
% Adaptation to almost zero. Checkpoint paths wrap at their punctuation.
\begin{tabularx}{\textwidth}{@{}
>{\raggedright\arraybackslash}p{2.20cm}
>{\raggedright\arraybackslash}p{1.90cm}
>{\raggedright\arraybackslash}p{2.70cm}
>{\raggedright\arraybackslash}p{2.00cm}
>{\raggedright\arraybackslash}X
>{\raggedright\arraybackslash}p{1.90cm}@{}}
\toprule
\textbf{Evaluated model} &
\textbf{CLAP family} &
\textbf{Exact public ancestor} &
\textbf{Encoder reuse} &
\textbf{Adaptation} &
\textbf{Task} \\
\midrule

\multicolumn{6}{@{}l}{\textsc{Attack ensemble checkpoints}} \\
\cmidrule(lr){1-6}

\textbf{MS-CLAP-22}
    & Microsoft CLAP
    & \path{CLAP_weights_2022.pth}
    & Native audio/text encoders
    & None
    & Classification, retrieval \\
\addlinespace[3pt]

\textbf{MS-CLAP-23}
    & Microsoft CLAP
    & \path{CLAP_weights_2023.pth}
    & Native audio/text encoders
    & None
    & Classification, retrieval \\
\addlinespace[3pt]

\textbf{LAION-CLAP}
    & LAION-CLAP
    & \path{630k-audioset-best.pt}
    & Native audio/text encoders
    & None
    & Classification, retrieval \\
\addlinespace[3pt]

\textbf{LAION-CLAP\textsuperscript{*}}
    & LAION-CLAP
    & \path{630k-audioset-fusion-best.pt}
    & Native audio/text encoders
    & None
    & Classification, retrieval \\

\addlinespace[6pt]
\multicolumn{6}{@{}l}{\textsc{Downstream evaluation models}} \\
\cmidrule(lr){1-6}

\textbf{EnCLAP}
    & LAION-CLAP
    & \path{630k-audioset-fusion-best.pt}
    & Precomputed CLAP audio embeddings
    & Frozen
    & Captioning \\
\addlinespace[4pt]

\textbf{ClapCap}
    & Microsoft CLAP
    & \path{CLAP_weights_2023.pth}
    & MS-CLAP-23 audio encoder + mapper
    & Audio encoder and GPT-2 frozen; mapper trained
    & Captioning \\
\addlinespace[4pt]

\textbf{AF1}
    & Microsoft CLAP (ClapCap)
    & MS-CLAP-23 via ClapCap
    & ClapCap audio encoder
    & Encoder frozen; transformation and cross-attention trained
    & Classification, Retrieval, captioning  \\
\addlinespace[4pt]

\textbf{Pengi}
    & Custom MS-CLAP-style model
    & Fine-tuned and Adapted
    & CLAP-pretrained HTSAT
    & Full encoder fine-tuning; learned mappers
    & Classification, captioning  \\
\addlinespace[4pt]

\textbf{AF2}
    & Custom AF-CLAP
    & Custom NVIDIA AF-CLAP (no public ancestor)
    & AF-CLAP audio encoder
    & Staged fine-tuning; intermediate CLAP update
    & Classification, retrieval, captioning  \\

\bottomrule
\end{tabularx}
\end{table}

\noindent\textbf{Public CLAP checkpoints.}
We evaluate the four public checkpoints used by the attack ensemble: MS-CLAP-22, MS-CLAP-23, LAION-CLAP \texttt{630k-audioset-best.pt}, and LAION-CLAP \texttt{630k-audioset-fusion-best.pt}. These experiments establish attack success on the constituent public representations available to the attacker.

\noindent\textbf{EnCLAP.}
EnCLAP \cite{kim_enclap_2024} uses precomputed LAION-CLAP audio embeddings as one input to its captioning architecture. The CLAP embeddings remain frozen. EnCLAP additionally represents the audio through a neural audio codec, so its final caption depends on both CLAP-derived semantic information and a separate acoustic representation before downstream language-model decoding.

\noindent\textbf{ClapCap.}
ClapCap \cite{elizalde_natural_2024} uses the MS-CLAP-23 audio encoder together with a learned mapper and GPT-2. The CLAP audio encoder and GPT-2 remain frozen while the mapper is trained to transform the CLAP audio representation into the language model's input space.

\noindent\textbf{Audio Flamingo (AF1).}
AF1 \cite{kong_audio_2024} incorporates the ClapCap audio encoder as its audio front end. That encoder remains frozen while additional transformations and cross-attention are trained inside the larger audio-language architecture. Thus, the final model output is separated from the CLAP-derived representation by multiple learned downstream components.

\noindent\textbf{Pengi.}
Pengi \cite{deshmukh_pengi_2024} uses a CLAP-pretrained HTSAT audio encoder as its initialization. Unlike the frozen-encoder systems, Pengi subsequently fine-tunes the audio encoder itself and learns additional mappings connecting the audio representation to the rest of the model. The public CLAP representation therefore serves as an initialization rather than a fixed downstream feature.

\noindent\textbf{Audio Flamingo 2 (AF2).}
AF2 \cite{ghosh_audio_2025} uses a custom AF-CLAP encoder rather than reusing any attacked public checkpoint. The AF-CLAP encoder is independent of public CLAP checkpoint parameters and then further adapted during AF2's staged training procedure. AF2 therefore shares CLAP-family design ancestry with the attacked models but no attacked public checkpoint parameters.

\subsection{Dataset Details}
\label{app:dataset-details}

\noindent\textbf{ESC-50.}
ESC-50 is an environmental-sound classification benchmark containing 50 classes of everyday acoustic events \cite{piczak_esc_2015}. Its class structure permits targeted and untargeted classification attacks and lets us construct source and target concepts from multiple recordings of the same semantic class.

\noindent\textbf{UrbanSound8K.}
UrbanSound8K is an urban-sound classification dataset containing realistic environmental recordings and official cross-validation folds \cite{noauthor_dataset_nodate}. Its classes include security-relevant events such as gunshots and sirens.

\noindent\textbf{AudioCaps.}
AudioCaps contains diverse real-world recordings paired with natural-language captions \cite{kim_audiocaps_2019}. We use AudioCaps for audio-to-text retrieval, text-to-audio retrieval, and automated audio captioning.

\noindent\textbf{Clotho.}
Clotho contains diverse environmental recordings paired with multiple human-written captions \cite{drossos_clotho_2019}. We use the official test split for retrieval and captioning.

\subsection{Downstream Task Details}

\subsubsection{Zero-Shot Audio Classification}
We report four classification metrics on the same fixed targeted cohort.

\noindent\textbf{Top-1 accuracy ($\downarrow$) and Top-5 accuracy ($\downarrow$).}
The fraction of adversarial samples for which the correct source class is the model's highest-ranked prediction, or among the top $5$ highest-ranked predictions.
% \noindent\textbf{Top-5 accuracy ($\downarrow$).}
% The fraction of adversarial samples for which the correct source class remains among the five highest-ranked predictions.

\noindent\textbf{Untargeted ASR ($\uparrow$).}
The fraction of samples that are correctly classified before the attack and incorrectly classified afterward. We condition this metric on clean-correct examples so that an attack receives credit only for errors that it causes. Because \textsc{ConceptAttack} synthesizes new audio, we instead condition on the same seed without CLAP guided denoising being classified correctly. Therefore, wrongly classified samples from the \textsc{ConceptAttack}s only contribute to the Untargeted ASR when the sample is correctly classified with the same seeded generation.

\noindent\textbf{Targeted ASR ($\uparrow$).}
The fraction of adversarial samples for which the model's highest-ranked prediction is the attacker-chosen target class. Targeted ASR is unconditioned.

\subsection{Retrieval Evaluation}
\label{app:retrieval}

We evaluate retrieval in the shared audio--text embedding space of each model.
Audio and text embeddings are $\ell_2$-normalized, candidates are ranked by
descending cosine similarity, and the clean and adversarial evaluations use
the same query cohort and candidate gallery.  Importantly, the gallery is the
full evaluation gallery rather than only the attacked examples.  For datasets
with multiple captions per recording, all captions associated with the same
audio item are treated as positives; the rank of the highest-ranked positive
determines whether retrieval succeeds.
%
% For a query $i$, let $r_{i,c}$ denote the best relevant rank under condition
% $c\in\{\mathrm{clean},\mathrm{adv}\}$ and define
% $H_{i,c}^{K}=\mathbf{1}[r_{i,c}\leq K]$.
%

We report results at $K\in\{1,5,10\}$, matching the standard R@1/5/10 cutoffs used for
retrieval.  These cutoffs also represent the small retrieved sets commonly
consumed by retrieval-augmented audio systems such as RECAP
\cite{ghosh_recap_2024}.  The clean exact-seed control differs from its
adversarial counterpart only in whether the source waveform is perturbed.  We
use the same fixed targeted cohort for targeted and untargeted metrics; apart
from conditioning untargeted ASR on clean success, we do not select examples
based on their observed retrieval outcome.

\subsection{Audio-to-Text Retrieval}

In audio-to-text (A2T) retrieval, a clean or adversarial audio query is ranked
against the text gallery.  A source hit occurs when at least one caption
associated with the query's source recording appears in the top-$K$ results.

\noindent\textbf{Untargeted ASR@$K$ ($\uparrow$).}
The fraction of queries for which the clean exact-seed control retrieves the
source within the top-$K$, but the adversarial query does not.  We compute the
metric only over queries for which the clean control succeeds at the
corresponding cutoff, avoiding failures that were already present without the
attack.

\noindent\textbf{Targeted ASR@$K$ ($\uparrow$).}
The fraction of targeted queries for which the attack's designated target text
appears within the top-$K$ results after the attack.  This metric is
unconditional: a target already retrieved by the clean control remains a
post-attack target hit (this rarely occurs in practice).

\subsection{Text-to-Audio Retrieval}

In text-to-audio (T2A) retrieval, all text queries remain benign and
unmodified.  The adversary controls only the audio items inserted into the
retrieval corpus.  For each query, the corresponding attacked item replaces
its clean exact-seed control while the rest of the gallery is held fixed.
Consequently, a targeted success means that attacker-chosen audio enters the
top-$K$ results returned for an ordinary text query.

\noindent\textbf{Untargeted ASR@$K$ ($\uparrow$).}
The fraction of benign text queries for which the associated clean audio is
retrieved within the top-$K$, but its adversarial replacement is not.  As in
A2T, we compute the metric only over queries for which the clean control
succeeds at the corresponding cutoff.

\noindent\textbf{Targeted ASR@$K$ ($\uparrow$).}
The fraction of targeted queries for which the attacker-chosen audio appears
within the top-$K$ results for its designated benign text query.  As in A2T,
this is the unconditional adversarial target-hit rate, and we use the same
fixed target assignments at all three cutoffs.

\subsection{Perturbation Attack Details}
\label{app:perturbation-attacks}

\paragraph{\textsc{clapPGD}}
We apply a targeted PGD attack with an equally weighted ensemble of four audio
encoders: LAION-CLAP with and without fusion, and Microsoft CLAP 2022 and 2023.
The target is represented by either a target-class
audio centroid, or a single audio embedding when there is only one target audio. Dropout is enabled during optimization, with new dropout masks
sampled at every iteration, but disabled when computing the clean and final
embeddings. We use 1000 iterations, random initialization, and constrain the
maximum waveform perturbation to \(\epsilon=0.0025\) unless otherwise mentioned. The source repulsion to targeted attraction is on a ratio of $2$ to $5$. 

\paragraph{Adversarial Illusions.}
This attack uses the same ensemble and PGD procedure, but represents the target
using its text embedding rather than an audio embedding. The target text is
either the selected class name or the target caption. Dropout is not enabled
for this attack, so the encoders remain deterministic during optimization.

\subsection{Average SNR of signals}
Table \ref{tab:attack-snr} shows the signal-to-noise ratio of the attacks for $\epsilon=\{0.001, 0.0025\}$.

\begin{table}[H]
    \centering
    \caption{Average signal-to-noise ratio (SNR) in dB.}
    \label{tab:attack-snr}
    \footnotesize
    \setlength{\tabcolsep}{2.5pt}
    \begin{tabular}{@{}lrrrr@{}}
        \toprule
        & \multicolumn{2}{c}{Adv. Illusion}
        & \multicolumn{2}{c}{\textsc{clapPGD}} \\
        \cmidrule(lr){2-3}
        \cmidrule(l){4-5}
        Dataset
        & $\epsilon=.001$
        & $\epsilon=.0025$
        & $\epsilon=.001$
        & $\epsilon=.0025$ \\
        \midrule
        ESC-50       & 38.50 & 31.64 & 38.50 & 31.60 \\
        UrbanSound8K & 35.14 & 27.99 & 35.21 & 28.07 \\
        \bottomrule
    \end{tabular}
\end{table}

\FloatBarrier
% The first results table occupies almost a full page; keep its section
% heading and introduction on that page as well.
\clearpage
\section{Full Experimental Results}
\label{app:fullexperimentalresults}
We now show the full results for \emph{Classification} in Table \ref{tab:zsc-classification}, \emph{Audio-to-text Retrieval} in Table \ref{tab:audio_to_text_retrieval_asr}, and  \emph{Text-to-audio Retrieval} in Table \ref{tab:text_to_audio_retrieval_asr}.

\begin{table}[H]
\centering
\caption{Results for zero-shot audio classification for the CLAP checkpoints and downstream task.}
\label{tab:zsc-classification}
\newcommand{\ZSCCleanLabel}{\textcolor{black!55}{(Clean)}}
\newcommand{\ZSCCleanValue}[1]{\textcolor{black!55}{(#1)}}
\scriptsize
\setlength{\tabcolsep}{2.6pt}
\renewcommand{\arraystretch}{1.02}
\begin{tabular*}{\textwidth}{@{\extracolsep{\fill}}llllcccc@{}}
\toprule
\textbf{Dataset} &
&
\textbf{Model} &
\textbf{Attack} &
\textbf{Top-1 Acc. $\downarrow$} &
\textbf{Top-5 Acc. $\downarrow$} &
\textbf{Untargeted ASR $\uparrow$} &
\textbf{Targeted ASR $\uparrow$} \\
\midrule

\multirow{28}{*}{\rotatebox[origin=c]{90}{\textbf{ESC-50}}}
& \multirow{16}{*}{\rotatebox[origin=c]{90}{\textbf{Attacked Ensemble}}}
& \multirow{4}{*}{\textbf{MS-CLAP-22}}
& \ZSCCleanLabel & \ZSCCleanValue{.838} & \ZSCCleanValue{.960} & \multicolumn{1}{c}{\textemdash} & \multicolumn{1}{c}{\textemdash} \\
& & & Adv.\ Illusions & .050 & .130 & .940 & \textbf{.843} \\
& & & \textsc{clapPGD} & .033 & \textbf{.100} & .961 & .823 \\
& & & \textsc{ConceptAttack} & \textbf{.022} & .248 & \textbf{.963} & .768 \\
\cmidrule(lr){3-8}\addlinespace[0.10em]
& & \multirow{4}{*}{\textbf{MS-CLAP-23}}
& \ZSCCleanLabel & \ZSCCleanValue{.933} & \ZSCCleanValue{.998} & \multicolumn{1}{c}{\textemdash} & \multicolumn{1}{c}{\textemdash} \\
& & & Adv.\ Illusions & .018 & \textbf{.065} & .981 & .895 \\
& & & \textsc{clapPGD} & .023 & .098 & .976 & .908 \\
& & & \textsc{ConceptAttack} & \textbf{.002} & .194 & \textbf{.996} & \textbf{.918} \\
\cmidrule(lr){3-8}\addlinespace[0.10em]
& & \multirow{4}{*}{\textbf{LAION-CLAP}}
& \ZSCCleanLabel & \ZSCCleanValue{.900} & \ZSCCleanValue{.998} & \multicolumn{1}{c}{\textemdash} & \multicolumn{1}{c}{\textemdash} \\
& & & Adv.\ Illusions & .020 & .073 & \textbf{.981} & .808 \\
& & & \textsc{clapPGD} & .023 & \textbf{.070} & .975 & \textbf{.898} \\
& & & \textsc{ConceptAttack} & \textbf{.016} & .226 & .979 & .864 \\
\cmidrule(lr){3-8}\addlinespace[0.10em]
& & \multirow{4}{*}{\textbf{LAION-CLAP\textsuperscript{*}}}
& \ZSCCleanLabel & \ZSCCleanValue{.840} & \ZSCCleanValue{.943} & \multicolumn{1}{c}{\textemdash} & \multicolumn{1}{c}{\textemdash} \\
& & & Adv.\ Illusions & .015 & .078 & \textbf{.982} & \textbf{.968} \\
& & & \textsc{clapPGD} & .023 & \textbf{.063} & .973 & .878 \\
& & & \textsc{ConceptAttack} & \textbf{.012} & .234 & .980 & .882 \\
\cmidrule(lr){2-8}\addlinespace[0.15em]
& \multirow{12}{*}{\rotatebox[origin=c]{90}{\textbf{Downstream}}}
& \multirow{4}{*}{\DownstreamModelCell{AF1}{Frozen}}
& \ZSCCleanLabel & \ZSCCleanValue{.625} & \ZSCCleanValue{.820} & \multicolumn{1}{c}{\textemdash} & \multicolumn{1}{c}{\textemdash} \\
& & & Adv.\ Illusions & .393 & .555 & .448 & .001 \\
& & & \textsc{clapPGD} & .198 & .355 & .712 & .265 \\
& & & \textsc{ConceptAttack} & \textbf{.052} & \textbf{.182} & \textbf{.880} & \textbf{.454} \\
\cmidrule(lr){3-8}\addlinespace[0.10em]
& & \multirow{4}{*}{\DownstreamModelCell{Pengi}{Fine-tuned}}
& \ZSCCleanLabel & \ZSCCleanValue{.763} & \ZSCCleanValue{.913} & \multicolumn{1}{c}{\textemdash} & \multicolumn{1}{c}{\textemdash} \\
& & & Adv.\ Illusions & .573 & .768 & .302 & .028 \\
& & & \textsc{clapPGD} & .415 & .595 & .498 & .143 \\
& & & \textsc{ConceptAttack} & \textbf{.066} & \textbf{.172} & \textbf{.862} & \textbf{.318} \\
\cmidrule(lr){3-8}\addlinespace[0.10em]
& & \multirow{4}{*}{\DownstreamModelCell{AF2}{Independent}}
& \ZSCCleanLabel & \ZSCCleanValue{.493} & \ZSCCleanValue{.625} & \multicolumn{1}{c}{\textemdash} & \multicolumn{1}{c}{\textemdash} \\
& & & Adv.\ Illusions & .458 & .615 & .086 & .013 \\
& & & \textsc{clapPGD} & .410 & .598 & .183 & .033 \\
& & & \textsc{ConceptAttack} & \textbf{.080} & \textbf{.156} & \textbf{.692} & \textbf{.082} \\

\addlinespace[0.20em]
\midrule
\addlinespace[0.10em]

\multirow{28}{*}{\rotatebox[origin=c]{90}{\textbf{UrbanSound8K}}}
& \multirow{16}{*}{\rotatebox[origin=c]{90}{\textbf{Attacked Ensemble}}}
& \multirow{4}{*}{\textbf{MS-CLAP-22}}
& \ZSCCleanLabel & \ZSCCleanValue{.793} & \ZSCCleanValue{.968} & \multicolumn{1}{c}{\textemdash} & \multicolumn{1}{c}{\textemdash} \\
& & & Adv.\ Illusions & .148 & .602 & .816 & \textbf{.805} \\
& & & \textsc{clapPGD} & .102 & \textbf{.532} & \textbf{.874} & .803 \\
& & & \textsc{ConceptAttack} & \textbf{.089} & .756 & .851 & .700 \\
\cmidrule(lr){3-8}\addlinespace[0.10em]
& & \multirow{4}{*}{\textbf{MS-CLAP-23}}
& \ZSCCleanLabel & \ZSCCleanValue{.830} & \ZSCCleanValue{.981} & \multicolumn{1}{c}{\textemdash} & \multicolumn{1}{c}{\textemdash} \\
& & & Adv.\ Illusions & .149 & \textbf{.502} & .822 & .799 \\
& & & \textsc{clapPGD} & .146 & .564 & .832 & .817 \\
& & & \textsc{ConceptAttack} & \textbf{.022} & .744 & \textbf{.962} & \textbf{.889} \\
\cmidrule(lr){3-8}\addlinespace[0.10em]
& & \multirow{4}{*}{\textbf{LAION-CLAP}}
& \ZSCCleanLabel & \ZSCCleanValue{.823} & \ZSCCleanValue{.973} & \multicolumn{1}{c}{\textemdash} & \multicolumn{1}{c}{\textemdash} \\
& & & Adv.\ Illusions & .145 & .534 & .824 & .797 \\
& & & \textsc{clapPGD} & .143 & \textbf{.505} & .830 & \textbf{.839} \\
& & & \textsc{ConceptAttack} & \textbf{.100} & .722 & \textbf{.855} & .789 \\
\cmidrule(lr){3-8}\addlinespace[0.10em]
& & \multirow{4}{*}{\textbf{LAION-CLAP\textsuperscript{*}}}
& \ZSCCleanLabel & \ZSCCleanValue{.748} & \ZSCCleanValue{.967} & \multicolumn{1}{c}{\textemdash} & \multicolumn{1}{c}{\textemdash} \\
& & & Adv.\ Illusions & .130 & .460 & .827 & .830 \\
& & & \textsc{clapPGD} & .131 & \textbf{.453} & .826 & .775 \\
& & & \textsc{ConceptAttack} & \textbf{.056} & .678 & \textbf{.936} & \textbf{.856} \\
\cmidrule(lr){2-8}\addlinespace[0.15em]
& \multirow{12}{*}{\rotatebox[origin=c]{90}{\textbf{Downstream}}}
& \multirow{4}{*}{\DownstreamModelCell{AF1}{Frozen}}
& \ZSCCleanLabel & \ZSCCleanValue{.566} & \ZSCCleanValue{.835} & \multicolumn{1}{c}{\textemdash} & \multicolumn{1}{c}{\textemdash} \\
& & & Adv.\ Illusions & .425 & .761 & .382 & .127 \\
& & & \textsc{clapPGD} & .319 & .666 & .549 & .288 \\
& & & \textsc{ConceptAttack} & \textbf{.122} & \textbf{.522} & \textbf{.733} & \textbf{.356} \\
\cmidrule(lr){3-8}\addlinespace[0.10em]
& & \multirow{4}{*}{\DownstreamModelCell{Pengi}{Fine-tuned}}
& \ZSCCleanLabel & \ZSCCleanValue{.627} & \ZSCCleanValue{.894} & \multicolumn{1}{c}{\textemdash} & \multicolumn{1}{c}{\textemdash} \\
& & & Adv.\ Illusions & .473 & .827 & .354 & .081 \\
& & & \textsc{clapPGD} & .368 & .753 & .505 & .199 \\
& & & \textsc{ConceptAttack} & \textbf{.122} & \textbf{.578} & \textbf{.719} & \textbf{.278} \\
\cmidrule(lr){3-8}\addlinespace[0.10em]
& & \multirow{4}{*}{\DownstreamModelCell{AF2}{Independent}}
& \ZSCCleanLabel & \ZSCCleanValue{.399} & \ZSCCleanValue{.660} & \multicolumn{1}{c}{\textemdash} & \multicolumn{1}{c}{\textemdash} \\
& & & Adv.\ Illusions & .368 & .652 & .096 & .071 \\
& & & \textsc{clapPGD} & .326 & \textbf{.639} & .191 & .131 \\
& & & \textsc{ConceptAttack} & \textbf{.222} & .678 & \textbf{.609} & \textbf{.311} \\

\bottomrule
\end{tabular*}
\vspace{2pt}

{\textsuperscript{*} \small{Fusion checkpoint; unmarked LAION-CLAP denotes the non-fusion checkpoint.}}
\end{table}

\begin{table}[H]
\centering
\caption{Attack success rate for audio-to-text retrieval on AudioCaps and
Clotho. Both targeted and untargeted metrics use the same targeted cohort;
untargeted ASR@K is source-rank escape conditioned on the exact-seed clean
control retrieving the source within the top K.}

\small
\setlength{\tabcolsep}{2.8pt}
\renewcommand{\arraystretch}{1.08}
\begin{tabular*}{\textwidth}{@{\extracolsep{\fill}}llllcccccc@{}}
\toprule
& & & &
\multicolumn{3}{c}{\textbf{Untargeted ASR $\uparrow$}} &
\multicolumn{3}{c}{\textbf{Targeted ASR $\uparrow$}} \\
\cmidrule(lr){5-7}
\cmidrule(lr){8-10}
\textbf{Dataset} &
&
\textbf{Model} &
\textbf{Attack} &
\textbf{ASR@1} &
\textbf{ASR@5} &
\textbf{ASR@10} &
\textbf{ASR@1} &
\textbf{ASR@5} &
\textbf{ASR@10} \\
\midrule

\multirow{15}{*}[-19.229pt]{\rotatebox[origin=c]{90}{\footnotesize\bfseries AudioCaps}}

& \multirow[c]{12}{*}[-8.718pt]{\rotatebox[origin=c]{90}{\footnotesize\bfseries Attacked Ensemble}}
& \multirow{3}{*}{\textbf{MS-CLAP-22}}
& Adversarial Illusions & .965 & .961 & .942 & \textbf{.676} & \textbf{.790} & \textbf{.831} \\
& & & \textsc{clapPGD} & .994 & .986 & .970 & .085 & .267 & .394 \\

& & & \textsc{ConceptAttack} & \textbf{1.00} & \textbf{1.00} & \textbf{.989} & .412 & .682 & .775 \\
\cmidrule(lr){3-10}\addlinespace[0.15em]

& & \multirow{3}{*}{\textbf{MS-CLAP-23}}
& Adversarial Illusions & .997 & .989 & \textbf{.985} & \textbf{.881} & \textbf{.924} & \textbf{.942} \\
& & & \textsc{clapPGD} & \textbf{1.00} & \textbf{.997} & .984 & .141 & .338 & .450 \\

& & & \textsc{ConceptAttack} & \textbf{1.00} & .989 & \textbf{.985} & .401 & .685 & .789 \\
\cmidrule(lr){3-10}\addlinespace[0.15em]

& & \multirow{3}{*}{\textbf{LAION-CLAP}}
& Adversarial Illusions & .996 & .983 & .976 & \textbf{.928} & \textbf{.958} & \textbf{.964} \\
& & & \textsc{clapPGD} & .998 & .986 & .977 & .224 & .485 & .603 \\

& & & \textsc{ConceptAttack} & \textbf{1.00} & \textbf{1.00} & \textbf{.993} & .579 & .837 & .898 \\
\cmidrule(lr){3-10}\addlinespace[0.15em]

& & \multirow{3}{*}{\textbf{LAION-CLAP}\textsuperscript{*}}
& Adversarial Illusions & .998 & .993 & .981 & \textbf{.941} & \textbf{.965} & \textbf{.972} \\
& & & \textsc{clapPGD} & \textbf{1.00} & .990 & .980 & .237 & .484 & .620 \\
& & & \textsc{ConceptAttack} & \textbf{1.00} & \textbf{1.00} & \textbf{.991} & .639 & .866 & .920 \\
\cmidrule(lr){2-10}\addlinespace[0.15em]

& \multirow[c]{4.6}{*}[2.428pt]{\rotatebox[origin=c]{90}{\footnotesize\bfseries Downstream}}
& \multirow{3}{*}{\textbf{AF-CLAP}}
& Adversarial Illusions & .174 & .057 & .048 & .000 & .001 & .006 \\[0.45em]
& & & \textsc{clapPGD} & .229 & .078 & .077 & .000 & .001 & .006 \\[0.45em]

& & & \textsc{ConceptAttack} & \textbf{.972} & \textbf{.939} & \textbf{.910} & \textbf{.095} & \textbf{.230} & \textbf{.304} \\[0.45em]

\addlinespace[0.30em]
\midrule
\addlinespace[0.10em]

\multirow{15}{*}[-20.328pt]{\rotatebox[origin=c]{90}{\footnotesize\bfseries Clotho}}

& \multirow[c]{12}{*}[-9.668pt]{\rotatebox[origin=c]{90}{\footnotesize\bfseries Attacked Ensemble}}
& \multirow{3}{*}{\textbf{MS-CLAP-22}}
& Adversarial Illusions & \textbf{1.00} & .991 & .984 & \textbf{.662} & \textbf{.783} & .823 \\
& & & \textsc{clapPGD} & \textbf{1.00} & .991 & .986 & .062 & .152 & .231 \\
& & & \textsc{ConceptAttack} & \textbf{1.00} & \textbf{1.00} & \textbf{1.00} & .511 & .761 & \textbf{.824} \\
\cmidrule(lr){3-10}\addlinespace[0.15em]

& & \multirow{3}{*}{\textbf{MS-CLAP-23}}
& Adversarial Illusions & \textbf{1.00} & .997 & .983 & \textbf{.882} & \textbf{.922} & \textbf{.929} \\
& & & \textsc{clapPGD} & \textbf{1.00} & .995 & .979 & .027 & .068 & .104 \\
& & & \textsc{ConceptAttack} & \textbf{1.00} & \textbf{1.00} & \textbf{1.00} & .515 & .757 & .841 \\
\cmidrule(lr){3-10}\addlinespace[0.15em]

& & \multirow{3}{*}{\textbf{LAION-CLAP}}
& Adversarial Illusions & \textbf{1.00} & \textbf{1.00} & \textbf{1.00} & \textbf{.889} & \textbf{.938} & \textbf{.940} \\
& & & \textsc{clapPGD} & \textbf{1.00} & \textbf{1.00} & \textbf{1.00} & .025 & .064 & .112 \\

& & & \textsc{ConceptAttack} & .955 & .974 & .983 & .552 & .774 & .851 \\
\cmidrule(lr){3-10}\addlinespace[0.15em]

& & \multirow{3}{*}{\textbf{LAION-CLAP}\textsuperscript{*}}
& Adversarial Illusions & \textbf{1.00} & \textbf{1.00} & \textbf{1.00} & \textbf{.900} & \textbf{.944} & \textbf{.948} \\
& & & \textsc{clapPGD} & \textbf{1.00} & \textbf{1.00} & \textbf{1.00} & .031 & .061 & .084 \\
& & & \textsc{ConceptAttack} & \textbf{1.00} & \textbf{1.00} & .991 & .589 & .796 & .852 \\
\cmidrule(lr){2-10}\addlinespace[0.15em]

& \multirow[c]{4.6}{*}[2.278pt]{\rotatebox[origin=c]{90}{\footnotesize\bfseries Downstream}}
& \multirow{3}{*}{\textbf{AF-CLAP}}
& Adversarial Illusions & .271 & .178 & .117 & .000 & .000 & .000 \\[0.45em]
& & & \textsc{clapPGD} & .308 & .208 & .134 &  .000 & .000 & .000 \\[0.45em]

& & & \textsc{ConceptAttack} & \textbf{1.00} & \textbf{.976} & \textbf{.985} & \textbf{.031} & \textbf{.091} & \textbf{.146} \\[0.45em]

\addlinespace[3pt]
\bottomrule
\end{tabular*}
\vspace{2pt}

{\normalsize \textsuperscript{*} Fusion checkpoint; unmarked LAION-CLAP denotes the non-fusion checkpoint.}
\label{tab:audio_to_text_retrieval_asr}
\end{table}

\begin{table}[H]
\centering
\caption{Attack success rate for text-to-audio retrieval on AudioCaps and
Clotho. Both targeted and untargeted metrics use the same targeted cohort;
untargeted ASR@K is source-rank escape conditioned on the exact-seed clean
control retrieving the source within the top K.}
\small
\setlength{\tabcolsep}{2.8pt}
\renewcommand{\arraystretch}{1.08}
\begin{tabular*}{\textwidth}{@{\extracolsep{\fill}}llllcccccc@{}}
\toprule
& & & &
\multicolumn{3}{c}{\textbf{Untargeted ASR $\uparrow$}} &
\multicolumn{3}{c}{\textbf{Targeted ASR $\uparrow$}} \\
\cmidrule(lr){5-7}
\cmidrule(lr){8-10}
\textbf{Dataset} &
&
\textbf{Model} &
\textbf{Attack} &
\textbf{ASR@1} &
\textbf{ASR@5} &
\textbf{ASR@10} &
\textbf{ASR@1} &
\textbf{ASR@5} &
\textbf{ASR@10} \\
\midrule

\multirow{15}{*}[-19.229pt]{\rotatebox[origin=c]{90}{\footnotesize\bfseries AudioCaps}}

& \multirow[c]{12}{*}[-8.718pt]{\rotatebox[origin=c]{90}{\footnotesize\bfseries Attacked Ensemble}}
& \multirow{3}{*}{\textbf{MS-CLAP-22}}
& Adversarial Illusions & .964 & .907 & .847 & \textbf{.824} & .894 & .917 \\
& & & \textsc{clapPGD} & .995 & \textbf{.987} & \textbf{.969} & .224 & .517 & .653 \\
& & & \textsc{ConceptAttack} & \textbf{1.00} & .947 & .853 & .680 & \textbf{.904} & \textbf{.938} \\
\cmidrule(lr){3-10}\addlinespace[0.15em]

& & \multirow{3}{*}{\textbf{MS-CLAP-23}}
& Adversarial Illusions & .888 & .780 & .665 & \textbf{.980} & \textbf{.993} & \textbf{.995} \\
& & & \textsc{clapPGD} & \textbf{.999} & \textbf{.989} & \textbf{.974} & .174 & .518 & .665 \\
& & & \textsc{ConceptAttack} & .783 & .734 & .660 & .890 & .969 & .979 \\
\cmidrule(lr){3-10}\addlinespace[0.15em]

& & \multirow{3}{*}{\textbf{LAION-CLAP}}
& Adversarial Illusions & .986 & .976 & .968 & \textbf{.958} & \textbf{.975} & \textbf{.981} \\
& & & \textsc{clapPGD} & .987 & .984 & \textbf{.971} & .422 & .741 & .866 \\
& & & \textsc{ConceptAttack} & \textbf{1.00} & \textbf{1.00} & .900 & .717 & .893 & .934 \\
\cmidrule(lr){3-10}\addlinespace[0.15em]

& & \multirow{3}{*}{\textbf{LAION-CLAP}\textsuperscript{*}}
& Adversarial Illusions & .996 & .985 & .974 & \textbf{.965} & \textbf{.978} & \textbf{.982} \\
& & & \textsc{clapPGD} & .999 & \textbf{.987} & \textbf{.981} & .408 & .747 & .860 \\
& & & \textsc{ConceptAttack} & \textbf{1.00} & .972 & .899 & .779 & .911 & .950 \\
\cmidrule(lr){2-10}\addlinespace[0.15em]

& \multirow[c]{4.6}{*}[2.428pt]{\rotatebox[origin=c]{90}{\footnotesize\bfseries Downstream}}
& \multirow{3}{*}{\textbf{AF-CLAP}}
& Adversarial Illusions & .064 & .040 & .015 & .003 & .016 & .038 \\[0.45em]
& & & \textsc{clapPGD} & .099 & .073 & .036 & .002 & .030 & .061 \\[0.45em]
& & & \textsc{ConceptAttack} & \textbf{.917} & \textbf{.750} & \textbf{.766} & \textbf{.108} & \textbf{.371} & \textbf{.520} \\[0.45em]

\addlinespace[0.30em]
\midrule
\addlinespace[0.10em]

% & \multirow{3}{*}{\textbf{Pengi}}
% & Adversarial Illusions & & & & & & \\
% & & \textsc{clapPGD} & & & & & & \\
% & & \textsc{ConceptAttack} & & & & & & \\

% \midrule

\multirow{15}{*}[-20.328pt]{\rotatebox[origin=c]{90}{\footnotesize\bfseries Clotho}}

& \multirow[c]{12}{*}[-9.668pt]{\rotatebox[origin=c]{90}{\footnotesize\bfseries Attacked Ensemble}}
& \multirow{3}{*}{\textbf{MS-CLAP-22}}
& Adversarial Illusions & .997 & .988 & .985 & \textbf{.765} & \textbf{.839} & .863 \\
& & & \textsc{clapPGD} & .994 & .987 & .985 & .090 & .260 & .391 \\
& & & \textsc{ConceptAttack} & \textbf{1.00} & \textbf{.995} & \textbf{.986} & .603 & .825 & \textbf{.889} \\
\cmidrule(lr){3-10}\addlinespace[0.15em]

& & \multirow{3}{*}{\textbf{MS-CLAP-23}}
& Adversarial Illusions & \textbf{1.00} & .984 & .982 & \textbf{.949} & \textbf{.962} & \textbf{.971} \\
& & & \textsc{clapPGD} & .998 & \textbf{.985} & \textbf{.983} & .040 & .104 & .173 \\
& & & \textsc{ConceptAttack} & \textbf{1.00} & .950 & .919 & .826 & .925 & .955 \\
\cmidrule(lr){3-10}\addlinespace[0.15em]

& & \multirow{3}{*}{\textbf{LAION-CLAP}}
& Adversarial Illusions & .975 & \textbf{.976} & \textbf{.977} & \textbf{.946} & \textbf{.960} & \textbf{.963} \\
& & & \textsc{clapPGD} & \textbf{.985} & .973 & .969 & .041 & .144 & .213 \\
& & & \textsc{ConceptAttack} & .818 & .946 & .967 & .704 & .872 & .912 \\
\cmidrule(lr){3-10}\addlinespace[0.15em]

& & \multirow{3}{*}{\textbf{LAION-CLAP}\textsuperscript{*}}
& Adversarial Illusions & .971 & .970 & .955 & \textbf{.959} & .967 & .972 \\
& & & \textsc{clapPGD} & \textbf{1.00} & \textbf{.998} & \textbf{.998} & .046 & .127 & .179 \\
& & & \textsc{ConceptAttack} & .956 & .842 & .781 & .921 & \textbf{.974} & \textbf{.986} \\
\cmidrule(lr){2-10}\addlinespace[0.15em]

& \multirow[c]{4.6}{*}[2.278pt]{\rotatebox[origin=c]{90}{\footnotesize\bfseries Downstream}}
& \multirow{3}{*}{\textbf{AF-CLAP}}
& Adversarial Illusions & .135 & .093 & .090 & .000 & .001 & .005 \\[0.45em]
& & & \textsc{clapPGD} & .163 & .113 & .101 & .000 & .001 & .004 \\[0.45em]
& & & \textsc{ConceptAttack} & \textbf{1.00} & \textbf{1.00} & \textbf{1.00} & \textbf{.025} & \textbf{.066} & \textbf{.089} \\[0.45em]
% \cmidrule(lr){2-9}\addlinespace[0.15em]

% & \multirow{3}{*}{\textbf{Pengi}}
% & Adversarial Illusions & & & & & & \\
% & & \textsc{clapPGD} & & & & & & \\
% & & \textsc{ConceptAttack} & & & & & & \\

\addlinespace[3pt]
\bottomrule
\end{tabular*}
\vspace{2pt}

{\normalsize \textsuperscript{*} Fusion checkpoint; unmarked LAION-CLAP denotes the non-fusion checkpoint.}
\label{tab:text_to_audio_retrieval_asr}
\end{table}

\FloatBarrier
\section{Robustness to Downstream Fine-Tuning}
\label{app:finetuningnotadefense}
A reasonable defense might be to fine-tune the CLAP model used in the deployed system. To test whether standard fine-tuning reliably decreases attack vulnerability, we adapt the supervised AudioCaps fine-tuning protocol of AudioSetCaps \cite{bai_audiosetcaps_2025} to the publicly available non-fusion LAION-CLAP checkpoint, fine-tuned independently under five random seeds. We retain its 20-epoch schedule, \(10^{-5}\) learning rate, global batch size of 196, and symmetric contrastive objective. We then evaluate \textsc{ConceptAttack} and PGD with \(\epsilon\in\{10^{-5},0.0005,0.001,0.00125,0.0025,0.005,0.01\}\).

\textsc{ConceptAttack} targeted ASR decreases from 86.4\% to 82.1\% across 20 epochs of fine-tuning.

\begin{figure}[!htbp]
    \centering
    \includegraphics[width=0.55\linewidth]{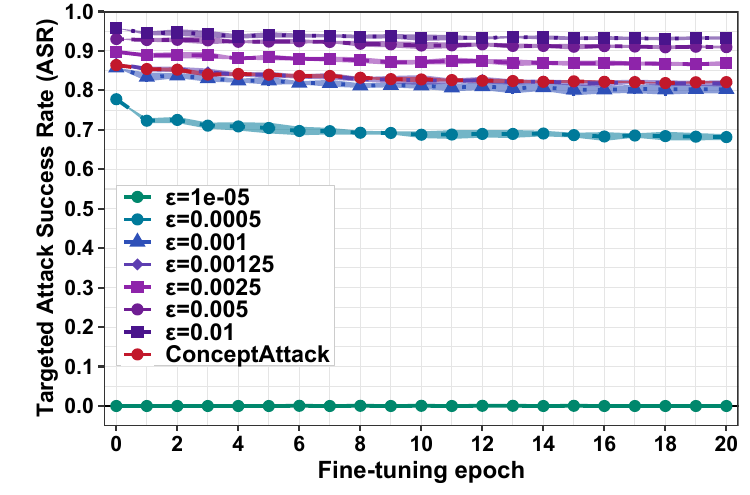}
    \setlength{\abovecaptionskip}{6pt}
    \caption{Targeted attack success rate (ASR) on ESC-50 across 20 epochs of AudioCaps fine-tuning for different PGD perturbation budgets ($\epsilon$). Lines show the mean over five seeded independently fine-tuned LAION-CLAP models, with shaded 95\% confidence intervals. Larger perturbation budgets consistently yield higher targeted ASR, and the light fine-tuning yields only a slight drop in ASR.}
    \label{fig:finetuning}
\end{figure}

\FloatBarrier
\section{Effect of Adversarial CLAP Guidance on Concept Attack}
\label{app:guidance-ablation}

We demonstrate the effect of adversarial CLAP guidance on \textsc{ConceptAttack} using zero-shot classification, one of the downstream tasks in our evaluation. As shown in Table~\ref{tab:zsc-guidance-ablation}, without CLAP guidance the same concept-generation pipeline almost never reaches the attacker-chosen target: targeted ASR is at most 1.6\% on ESC-50 and 7.8\% on UrbanSound8K across all evaluated models. With CLAP guidance, targeted ASR rises to 70.0--91.8\% on the attacked checkpoints and transfers downstream, reaching 45.4\% for Audio Flamingo 1 on ESC-50 and 35.6\% on UrbanSound8K. 

This provides strong evidence that synthesis alone does not produce the targeted attack success observed in \textsc{ConceptAttack}. Adversarial CLAP guidance drives the generated audio toward the attacker-chosen semantics, and that effect propagates into unseen downstream systems.

% Local one-line variant for this table only.
\newcommand{\InlineDownstreamModelCell}[2]{%
  {#1 (#2)}%
}

\begin{table}[H]
\centering
\setlength{\abovecaptionskip}{6pt}
\caption{Zero-shot audio classification with and without adversarial CLAP guidance.}
\label{tab:zsc-guidance-ablation}
\small
\begin{minipage}{\textwidth}
\centering
\setlength{\tabcolsep}{3pt}
\renewcommand{\arraystretch}{1.40}
\begin{tabularx}{\linewidth}{@{}ccl*{4}{>{\centering\arraybackslash}X}@{}}
\toprule
& & &
\multicolumn{2}{c}{\textbf{Without CLAP guidance}} &
\multicolumn{2}{c}{\textbf{With CLAP guidance}} \\
\cmidrule(lr){4-5}
\cmidrule(lr){6-7}
\textbf{Dataset} &
&
\textbf{Model} &
\textbf{Top-1 Acc. $\downarrow$} &
\textbf{Targeted ASR $\uparrow$} &
\textbf{Top-1 Acc. $\downarrow$} &
\textbf{Targeted ASR $\uparrow$} \\
\midrule

\multirow[c]{7}{*}[-5.023pt]{\rotatebox[origin=c]{90}{\normalfont ESC-50}}
& \multirow[c]{4}{*}[0.538pt]{\rotatebox[origin=c]{90}{\normalfont\shortstack[c]{Attacked\\Ensemble}}}
& MS-CLAP-22 & .700 & .004 & .022 & .768 \\
& & MS-CLAP-23 & .740 & .002 & .002 & .918 \\
& & LAION-CLAP & .620 & .006 & .016 & .864 \\
& & LAION-CLAP\textsuperscript{*} & .600 & .000 & .012 & .882 \\
\cmidrule(lr){2-7}
\addlinespace[0.3em]
& \multirow[c]{3.4}{*}[2.886pt]{\rotatebox[origin=c]{90}{\normalfont Downstream}}
& \InlineDownstreamModelCell{AF1}{Frozen} & .500 & .008 & .052 & .454 \\
& & \InlineDownstreamModelCell{Pengi}{Fine-tuned} & .560 & .012 & .066 & .318 \\
& & \InlineDownstreamModelCell{AF2}{Independent} & .180 & .016 & .080 & .082 \\

\addlinespace[0.4em]
\cmidrule(lr){2-7}
\addlinespace[0.4em]

\multirow[c]{7}{*}[-3.313pt]{\rotatebox[origin=c]{90}{\normalfont UrbanSound8K}}
& \multirow[c]{4}{*}[2.338pt]{\rotatebox[origin=c]{90}{\normalfont\shortstack[c]{Attacked\\Ensemble}}}
& MS-CLAP-22 & .600 & .044 & .089 & .700 \\
& & MS-CLAP-23 & .600 & .044 & .022 & .889 \\
& & LAION-CLAP & .700 & .033 & .100 & .789 \\
& & LAION-CLAP\textsuperscript{*} & .800 & .022 & .056 & .856 \\
\cmidrule(lr){2-7}
\addlinespace[0.3em]
& \multirow[c]{3.4}{*}[2.800pt]{\rotatebox[origin=c]{90}{\normalfont Downstream}}
& \InlineDownstreamModelCell{AF1}{Frozen} & .300 & .078 & .122 & .356 \\
& & \InlineDownstreamModelCell{Pengi}{Fine-tuned} & .600 & .044 & .122 & .278 \\
& & \InlineDownstreamModelCell{AF2}{Independent} & .500 & .056 & .222 & .311 \\

\addlinespace[0.42em]
\bottomrule
\end{tabularx}
\vspace{0.25em}

\parbox{\linewidth}{\centering
\textsuperscript{*}Fusion checkpoint; unmarked LAION-CLAP denotes the non-fusion checkpoint.\par}
\end{minipage}
\end{table}

\FloatBarrier
\section{Over-the-Air Experimental Setup}
\label{app:liverecording}

Our over-the-air experiments test whether the attacks survive playback and re-recording in notoriously acoustically imperfect environments, played through acoustically imperfect consumer-grade devices.

\noindent\textbf{Acoustic environments.}
We conduct the high-ambient-noise evaluation in an outdoor recording site in Hollywood's Vinyl (music recording) District on an arbitrarily chosen Friday evening (i.e., a busy time).

We conduct the in-vehicle evaluation inside an SUV, with the vehicle stationary, the engine off, and the windows closed. 

\noindent\textbf{Consumer playback chain.}
We render all source audio at 44.1~kHz and play it through popular (acoustically suboptimal) consumer-grade monitoring equipment: one studio monitor-speaker of an ADAM Audio D3V desktop monitoring system (which retails for approximately \$250 per pair). Audio is routed monophonically through the monitor; digital-to-analog conversion is performed by a MOTU M4 interface. For every recording, the capture device is positioned on the monitor's acoustic axis, with its microphone input 18~inches from the front baffle.

\noindent\textbf{Capture chain 1: Professional.}
For the professional capture condition, we use a Schoeps CMC-1 microphone amplifier with an MK5 small-diaphragm condenser capsule configured in cardioid mode for outdoor playback and omnidirectional mode for in-car playback, as is standard for location recording. We note that this microphone is widely regarded as the recording industry standard for accurate capture, so it reflects the upper bound of what can be recovered under these highly suboptimal playback locations and devices. We record uncompressed audio at 44.1~kHz (which matches the attack audio) and 32-bit floating-point resolution. We fit a standard Schoeps B5D windscreen to shield the MK5 capsule described herein from wind noise in the outdoor environment. We use no microphone wind protection foam for the vehicle environment, as is standard.

\noindent\textbf{Capture chain 2: Consumer phone (with resampling loss).}
For the consumer capture condition, we record using the built-in microphone array of an iPhone~16~Pro. We configure the application to record mono uncompressed audio at a 48~kHz sample rate, with all configurable on-device audio processing disabled. Recording at 48~kHz avoids sample-rate conversion during capture, as this iPhone is 48~kHz-native. Before evaluation, we resample the recordings to 44.1~kHz offline, using the identical processing step as described in our resampling robustness experiments.

% OLD WRITEUP AND NOTES
% \clearpage
% \input{Sections/OLD Introduction}
% \input{Sections/OLD Contribution}
% \input{Sections/OLD Threat-model}
% \input{Sections/OLD Methods}
% \input{Sections/OLD-ExperimentsSetup-AB}
% \input{Sections/OLD Results}
% \input{Sections/OLD Discussion}

\end{document}